\documentclass[useAMS,usenatbib]{mnras}
\usepackage{graphicx}
\usepackage{xcolor}
\usepackage{amssymb,amsmath,bm}
\usepackage{hyperref}
\usepackage[export]{adjustbox}
\hypersetup{colorlinks=true,citecolor=blue}
\usepackage{multirow}
\usepackage{atbegshi,picture}

\newcommand{\vb}[1]{\mathbf{#1}}
\def\ie{{\em i.e.}~}

\newcommand{\tj}[6]{ \begin{pmatrix}
   #1 & #2 & #3 \\
   #4 & #5 & #6 
  \end{pmatrix}}

\title[Wide angle effects for peculiar velocities]{Wide angle effects for peculiar velocities}

\author[Castorina \& White]
{Emanuele Castorina$^{1,2}$\thanks{e-mail: emanuele.castorina@cern.ch}, 
Martin White$^{2,3}$\thanks{e-mail: mwhite@berkeley.edu}
 \\~\\
\footnotesize
$^1$Theoretical Physics Department, CERN, 1211 Geneva 23, Switzerland\\
$^2$Berkeley Center for Cosmological Physics, University of California, Berkeley, CA 94720, USA\\
$^3$Lawrence Berkeley National Laboratory, 1 Cyclotron Road, Berkeley, CA 93720, USA\\
}

\AtBeginShipoutNext{\AtBeginShipoutUpperLeft{%
  \put(\dimexpr\paperwidth-1cm\relax,-1.cm){\makebox[0pt][r]{CERN-TH-2019-191}}%
}}
\begin{document}

\maketitle

\begin{abstract}
The line-of-sight peculiar velocities of galaxies contribute to their observed redshifts, breaking the translational invariance of galaxy clustering down to a rotational invariance around the observer.  This becomes important when the line-of-sight direction varies significantly across a survey, leading to what are known as `wide angle' effects in redshift space distortions.  Wide-angle effects will also be present in measurements of the momentum field, \ie the galaxy density-weighted velocity field, in upcoming peculiar velocity surveys. 
In this work we study how wide-angle effects modify the predicted correlation function and power spectrum for momentum statistics, both in auto-correlation and in cross-correlation with the density field.  Using both linear theory and the Zeldovich approximation, we find that deviations from the plane-parallel limit are large and could become important in data analysis for low redshift surveys. We point out that even multipoles in the cross-correlation between density and momentum are non-zero regardless of the choice of line of sight, and therefore contain new cosmological information that could be exploited.  We discuss configuration-space, Fourier-space and spherical analyses; providing exact expressions in each case rather than relying on an expansion in small angles.  We hope these expressions will be of use in the analysis of upcoming surveys for redshift-space distortions and peculiar velocities.
\end{abstract}

\section{Introduction}
\label{sec:intro}

Objects are not at rest in the expanding Universe, and the study of their peculiar motions provides us with an opportunity to test our models of gravity and structure formation as well as more tightly constrain the parameters of those models \citep{Weinberg13,Amendola18}.  The cosmic velocity field can be studied either indirectly through its impact on clustering statistics (so-called ``redshift-space distortions; \citealt{Kai87,H98,Pea99}) or directly by measuring the peculiar velocity field.  The latter, often called `cosmic flows' \citep{LavauxHudson,Tully16}, has a long history \citep{SW95}. Measurements of the (reconstructed) velocity field allowed different authors to place constraints on the growth of structure at low redshift \citep{Hudson12, Johnson14, Carrick15, Howlett17,Adams17, Feix17, Dupuy19, Qin19}, and several velocity surveys  are planned or in preparation \citep{Taipan,Wallaby,Kim19}. These surveys could measure the growth of structure at the few percent level at low redshift deep in dark energy dominated era, providing independent information from traditional galaxy surveys \citep{Koda2013,Howlett2017}.
In order to provide accurate measures of the velocity field both the redshift and distance of an object need to be known, and this limits velocity surveys to the relatively local Universe.  To obtain large volumes one is then forced to cover large areas of the sky.  This combination of low mean distance and large sky coverage means that for such surveys the plane-parallel limit, sometimes called distant-observer limit, usually employed when modeling the correlation function or power spectrum is a poor approximation. This becomes more and more of a problem for widely separated galaxies, hence the name of wide angle effects. Expressions for the velocity-velocity correlation function beyond the plane parallel limit are well known \citep{Gorski88,Nusser17}, while less attention has been devoted to wide angle effects in galaxy density-velocity cross correlations. In Fourier space , extra care is required when defining the velocity auto- and cross- power spectra on the curved sky. One of our goal is therefore to extend the treatment in \cite{CasWhi18a} of the density fields on the curved sky to the multipoles of the momentum (\ie density weighted galaxy velocities) auto and cross spectra.

The outline of the paper is as follows.  In \S\ref{sec:planeparallel} we define our notation and review the results in the distant observer or plane-parallel limit.
In \S\ref{sec:beyondPP} we present the full expression, in configuration space, beyond plane parallel for the mid-point and endpoint definition of the line-of-sight (LOS).  Section \ref{sec:fourier} discusses the Fourier space descriptions of the velocity fields, while \S\ref{sec:SFB} introduces the spherical-Fourier expansion.  We go beyond linear theory in \S\ref{sec:zeldovich}, where we discuss how to extend Lagrangian perturbation theory to the wide-angle regime.  We present our conclusions in \S\ref{sec:conclusions}.

\section{Momentum 2-point function in linear theory and the plane-parallel limit}
\label{sec:planeparallel}

We first introduce our notation and quickly review the main results in the plane-parallel or distant observer limit.  In this limit the observer is taken to be very far from the pair of points whose correlation we are investigating, which means that the line-of-sight direction can be taken to be the same for both points.  

Let us define the line-of-sight momentum as $\rho(\mathbf{s})=\left(1+\delta(\mathbf{s})\right)u(\mathbf{s})$, where $u(\mathbf{s})$ is the line-of-sight component of the velocity in Hubble units, i.e.~$u(\mathbf{s})=\hat{s}\cdot\mathbf{v}(\mathbf{s})/(aH)$, and $\mathbf{s}$ denotes the redshift-space position.  In linear theory the momentum is equal to the velocity, $u(\mathbf{s})$, and because there is no ``mean velocity'' in linear theory $\mathbf{u}$ is the same in real as in redshift space.

In linear theory the density and velocity are related via $\mathbf{v}(\mathbf{k})=i(aHf)(\mathbf{k}/k^2)\delta(\mathbf{k})$.  Thus in the plane-parallel or distant observer limit $u(\mathbf{k})=if(k_z/k^2)\delta(\mathbf{k})=(if\mu/k)\delta(\mathbf{k})$ with $\mu=\hat{k}\cdot\hat{z}$ and $\hat{z}$ the (common) line of sight. For example, the momentum auto-power spectrum (which equals the velocity auto-power spectrum in linear theory) is simply
\begin{equation}
    P^{\rho\rho}(k,\mu)=\frac{\mu^2f^2}{k^2}\ P(k)
\label{eqn:rhorho-pp}
\end{equation}
where $P(k)$ is the linear theory density power spectrum, which depends only upon $k=|\mathbf{k}|$.
This expression differs by a factor of $(aH)^2$ from the similar expression in \citet{Park00} due to our choice of units for $u(\mathbf{r})$.  As noted by \citet{Howlett19}, the momentum power spectrum is the same in real- and redshift-space in the limit of linear theory.
Assuming a scale-independent, linear bias ($b$) the momentum-density cross spectrum is
\begin{equation}
    P^{\rho\delta}(k,\mu)=\frac{if\mu}{k}P(k) \left( b + f\mu^2 \right)
\label{eqn:rhodelta-pp}
\end{equation}
and is pure imaginary.  This is required in order for the corresponding cross-spectrum in configuration space to be real, since velocities change sign under a parity transformation.

The relevant correlation functions are
\begin{align}
    \xi_{\delta\delta}(s,\mu_s) &= \left(b^2+\frac{2bf}{3}+\frac{f^2}{5}\right)\mathcal{L}_0\Xi_0^{(0)}
    \nonumber \\
    &- \left(\frac{4bf}{3}+\frac{4f^2}{7}\right)\mathcal{L}_2\Xi_2^{(0)} +
    \frac{8f^2}{35}\mathcal{L}_4\Xi_4^{(0)} \\
    \xi_{\delta\rho}(s,\mu_s) &= -f\left(b+\frac{3f}{5}\right)\mathcal{L}_1\Xi_1^{(1)}
    + \frac{2f^2}{5}\mathcal{L}_3\Xi_3^{(1)} \\
    \xi_{\rho\rho}(s,\mu_s) &= \frac{f^2}{3}\Xi_0^{(0)}\mathcal{L}_0
    -\frac{2f^2}{3}\mathcal{L}_2\Xi_2^{(2)}
\end{align}
where $\mathcal{L}_\ell$ is the Legendre polynomial of order $\ell$, with suppressed argument $\mu_s=\hat{s}\cdot\hat{z}$, and we have defined
\begin{equation}
    \Xi_\ell^{(n)}(s) = \int\frac{k^2\,dk}{2\pi^2}\ k^{-n}P(k)j_\ell(ks)
\end{equation}

\section{Beyond plane parallel}
\label{sec:beyondPP}

Once we drop the distant observer approximation there are two classes of effects to consider.  The first is the impact of a varying line-of-sight direction upon the dynamical fields.  The second is the change in number density and volume element with a radial distance change induced by redshift-space distortions and it's not a dynamical effect.  The latter leads to corrections going as $1/s$ and depending upon the mean density of the sample, $\bar{n}(\mathbf{s})$.  \citet{Sza98} parameterized these corrections as ``$\alpha$ terms'', and we shall follow this convention.  Since they do not play a large role in what follows, we shall relegate the presentation of the $\alpha$ terms to Appendix \ref{app:alpha}.

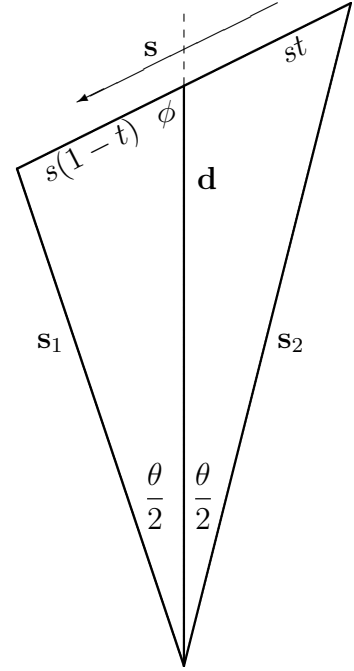
\begin{figure}
\begin{center}
\begin{picture}(260,260)
\Large
\thicklines
\qbezier(125.0,0.0)(125.0,112.5)(125.0,218.8)
\qbezier(125.0,0.0)(156.2,125.0)(187.5,250.0)
\qbezier(125.0,0.0)(93.8,93.8)(62.5,187.5)
\qbezier(62.5,187.5)(125.0,218.8)(187.5,250.0)
\thinlines
\multiput(125,219)(0,6){5}{\line(0,1){3}}
\put(110,60){$\displaystyle{\frac{\theta}{2}}$}
\put(128,60){$\displaystyle{\frac{\theta}{2}}$}
\put( 70,120){$\mathbf{s}_1$}
\put(160,120){$\mathbf{s}_2$}
\put(130,180){$\mathbf{d}$}
\put( 70,182){\rotatebox{27}{$s(1-t)$}}
\put(160,227){\rotatebox{27}{$st$}}
\put(115,205){$\phi$}
\put(160,250){\vector(-2,-1){75}}
\put(110,230){$\mathbf{s}$}
\end{picture}
\caption{The assumed geometry and angles.  The two galaxies lie at $\mathbf{s}_1$
and $\mathbf{s}_2$, with separation vector $\mathbf{s}=\mathbf{s}_1-\mathbf{s}_2$ and
form an angle $\theta$.
We take the line of sight to be either parallel to the angle bisector, $\mathbf{d}$,
which divides $\mathbf{s}$ into parts of lengths $st$ and $s(1-t)$ or to the direction
$\mathbf{s}_1$.
The separation vector between the two galaxies, $\mathbf{s}$, makes an angle $\phi$ with the line of sight
direction, $\hat{d}$.}
\label{fig:triangle}
\end{center}
\end{figure}

To extend the plane-parallel calculation to include wide angle effects we return to the linear theory, redshift-space density field which can be written \citep{Kai87}
\begin{equation}
    \delta^{(s)}(\mathbf{s}) = \int\frac{d^3k}{(2\pi)^3}
    \ e^{i\mathbf{k}\cdot\mathbf{s}}\left[b+f\left(\hat{k}\cdot\hat{s}\right)^2\right]
    \delta(\mathbf{k})
\label{eqn:deltas-defn}
\end{equation}
while the radial velocity field is
\begin{equation}
    u(\mathbf{s}) = f\int\frac{d^3k}{(2\pi)^3}
    \ e^{i\mathbf{k}\cdot\mathbf{s}}\frac{i\hat{k}\cdot\hat{s}}{k}
    \delta(\mathbf{k})
\end{equation}
Following \citet{Sza98,CasWhi18a} we decompose these into multipole moments
\begin{equation}
    \delta_\ell^n \equiv \int\frac{d^3k}{(2\pi)^3}\ e^{i\mathbf{k}\cdot\mathbf{s}}
    k^{-n}\mathcal{L}_\ell\left(\hat{k}\cdot\hat{s}\right)\delta(\mathbf{k})
    \label{eq:delta_ell}
\end{equation}
where $\mathcal{L}_\ell$ is the Legendre polynomial of order $\ell$.  It is then straightforward to compute
\begin{align}
    \left\langle\delta_{\ell_1}^{n_1}\delta_{\ell_2}^{n_2}\right\rangle &=
    \frac{(4\pi)^2}{(2\ell_1+1)(2\ell_2+1)}\int\frac{k^2\,dk}{2\pi^2}
    \frac{P(k)}{k^{n_1}k^{n_2}} \sum_L i^L j_L(ks) \nonumber \\
    &\times \sum_{m_1m_2M}\mathcal{G}_{Mm_1m_2}^{L\ell_1\ell_2} Y^\star_{LM}(\hat{s})Y^\star_{\ell_1 m_1}(\hat{s}_1)Y^\star_{\ell_2 m_2}(\hat{s}_2)
\label{eqn:deltadelta}
\end{align}
where $\mathcal{G}$ is the Gaunt integral and the sum only involves a few non-zero terms \citep[see also][for related derivations in the context of velocities]{Ma11,Nusser17}. Notice that Eqs. \ref{eq:delta_ell,eqn:deltadelta} remain valid beyond linear theory.
Unless otherwise noted in the remainder of this paper we will assume $b=1$.
Upon dropping the plane parallel assumption our two-point function becomes a function of `triangles' and we must define a line of sight and geometry for our pair.  There are two line-of-sight definitions which we will find useful.  The first is the angle bisector, which is shown as the vector $\mathbf{d}$ in Fig.~\ref{fig:triangle}.  We shall specify the triangle by giving $s$, $d$ and $\mu=\hat{s}\cdot\hat{d}$.  The second is to pick one of the directions, $\mathbf{s}_1$ or $\mathbf{s}_2$ in Fig.~\ref{fig:triangle}, as a line of sight.  This is particularly useful when computing power spectra as it allows a factorization of the calculation which dramatically improves efficiency.  We shall pick $\mathbf{s}_1$ as our line of sight and refer to this convention as the ``endpoint definition''.  We will specify the triangle by giving $s$, $s_1$ and $\mu_1=\hat{s}\cdot\hat{s}_1$.

We first give the results in configuration space for the bisector definition and then turn to a discussion of the endpoint results in configuration space.  These results will extend the earlier work of \citet{Sza98} and \citet{Rei16,CasWhi18a} to the velocity statistics.
In the next section (\S\ref{sec:fourier}) we will discuss the Fourier space statistics, which require some care in their definition.

\subsection{Angle bisector}

In the angle bisector parameterization we set the line of sight ($\hat{d}$) to be the $\hat{z}$ axis and orient the triangle in Fig.~\ref{fig:triangle} to lie in the $x-z$ plane so that all of the polar angles are zero or $\pi$.  Both $\hat{s}_1$ and $\hat{s}_2$ lie at $\theta/2$ to the $z$-axis while $\hat{s}$ is at $\pi-\phi$.  The contributions in Eq.~\ref{eqn:deltadelta} can then be calculated using the explicit expressions for $\mathcal{G}$ and $Y_{\ell m}$.

\begin{figure*}
    \centering
    \includegraphics[width=\textwidth]{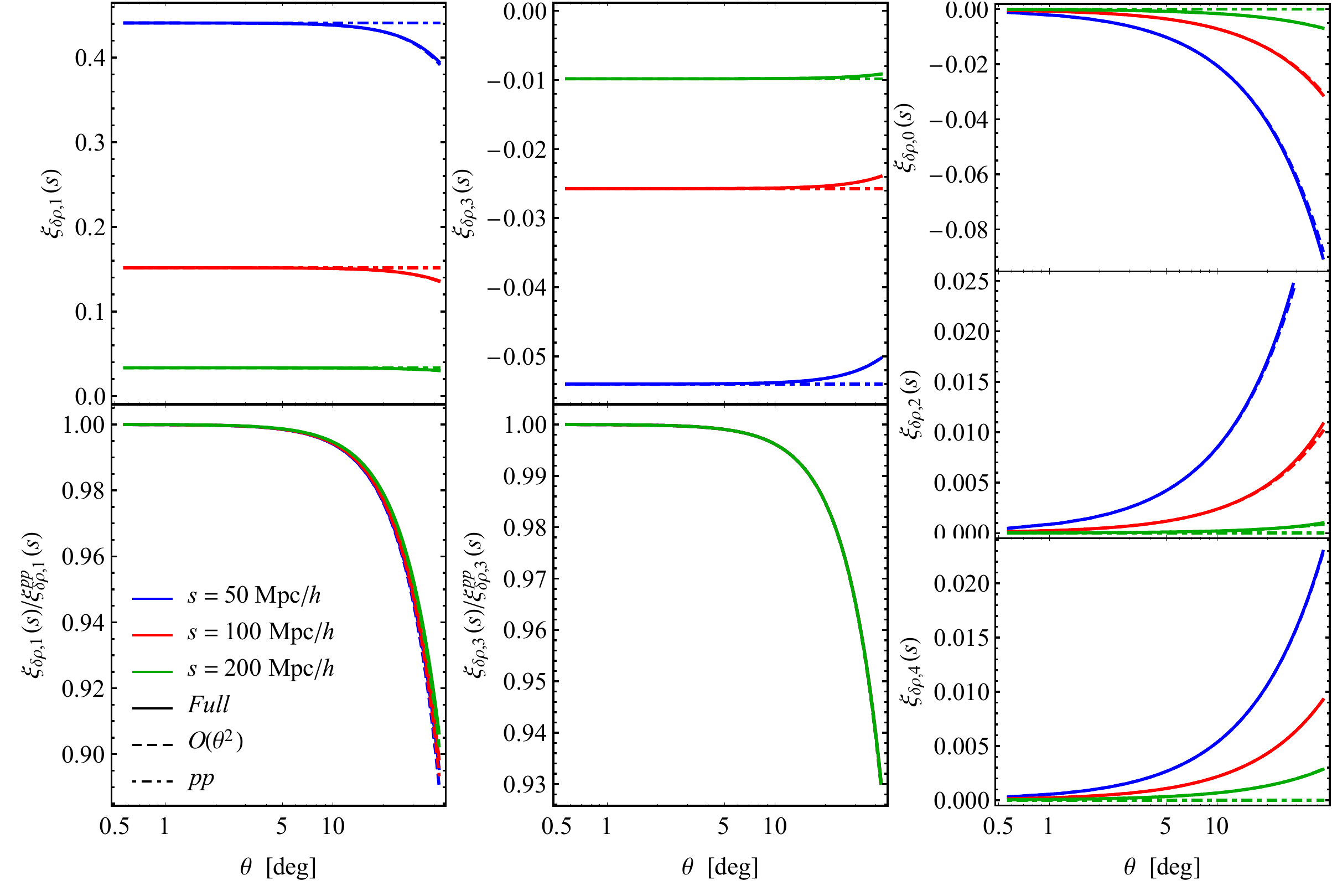}
    \caption{The multipoles of the density-momentum cross-correlation in the bisector parameterization for fixed separations as a function of opening angle, $\theta$.  We have assumed $b=1$.  The left two columns show the dipole and octopole, with the upper row showing the signal and the lower row the ratio to plane-parallel ($\theta\to 0$).  The right column shows the even multipoles, which are zero in the plane-parallel limit.}
\label{fig:fig_bis}
\end{figure*}

For the density-density auto-correlation the corrections can be found\footnote{Note that Eq.~(15) of \citet{Sza98} contains a typographical error.  The $4/15$ should be $8/15$.  Also beware that the $\theta$ of \citet{Sza98} is half our definition.} in \citet{Sza98,CasWhi18a} (see also Appendix \ref{app:density}).  For the velocity-velocity auto-correlation
\begin{align}
    \left\langle u(\mathbf{s}_1)u(\mathbf{s}_2)\right\rangle
    &= \frac{f^2}{3}\cos\theta\ \Xi_0^{(2)}(s)\mathcal{L}_0(\mu_s)
    \nonumber \\
    &- \frac{2f^2}{3}\left(\mathcal{L}_2(\mu_s)-\frac{1}{4}\left[1-\cos\theta\right]\right)\Xi_2^{(2)}(s)
\end{align}
We can expand this for small angles using the identities in Appendix A of \citet{CasWhi18a}
\begin{align}
    \left\langle u(\mathbf{s}_1)u(\mathbf{s}_2)\right\rangle_{\ell=0}
    &= \frac{f^2}{3}\left[ \Xi_0^{(2)}(s) - \frac{x^2}{3}\Xi_0^{(2)}(s) + \frac{x^2}{6}\Xi_2^{(2)}
     \right] + \cdots \\
    \left\langle u(\mathbf{s}_1)u(\mathbf{s}_2)\right\rangle_{\ell=2}
    &= \frac{f^2}{3}\left[
    -2\Xi_2^{(2)}(s)+\frac{x^2}{3}\Xi_0^{(2)}(s) - \frac{x^2}{6}\Xi_2^{(2)}(s) \right]
    + \cdots
\end{align}
where $x\equiv s/d\ll 1$ for small angles.  Note that for the bisector definition of the line of sight, and for the velocity auto-correlation function, the first correction is $\mathcal{O}(x^2)$.  Further note that when $\theta\ne 0$ the index of the $j_L$ in $\Xi_L^{(n)}$ and the index of $\mathcal{L}_\ell(\mu_s)$ need no longer match, which makes the Fourier expressions more complicated \citep{CasWhi18a}.  We shall address Fourier space in \S\ref{sec:fourier}.

For the cross term one finds
\begin{align}
    \left\langle \delta(\mathbf{s}_1)u(\mathbf{s}_2)\right\rangle
    &= -f \cos\frac{\theta}{2}\left(b-\frac{f}{5}+\frac{4f}{5}\cos^2\frac{\theta}{2}\right)
    \mathcal{L}_1(\mu_s)\Xi_1^{(1)}(s) \nonumber \\
    &+ f\sin\frac{\theta}{2}\left(b-\frac{f}{5}+\frac{4f}{5}\sin^2\frac{\theta}{2}\right)
    \sqrt{1-\mu_s^2}\,\Xi_1^{(1)}(s) \nonumber \\
    &+ \frac{2f^2}{5}\cos\frac{\theta}{2}\mathcal{L}_3(\mu_s)\Xi_3^{(1)}(s) \nonumber \\
    &+ \frac{f^2}{20}\left(\cos\frac{3\theta}{2}-\cos\frac{\theta}{2}\right)\mathcal{L}_1(\mu_s)\Xi_3^{(1)}(s)
    \nonumber \\
    &+ \frac{2f^2}{15}\sin\frac{\theta}{2}\left(5\mathcal{L}_2(\mu_s)+\frac{1+3\cos\theta}{4}\right)
    \nonumber \\
    &\times \sqrt{1-\mu_s^2}\,\Xi_3^{(1)}(s)
\end{align}
Unlike the $\delta\delta$ and $uu$ auto-correlations, the corrections to these spectra start at $\mathcal{O}(\theta)$, rather than $\mathcal{O}(\theta^2)$, even for the bisector definition of the line of sight.  Again expanding in powers of $x=s/d$ up to $x^2$:
\begin{align}
     \left\langle \delta(\mathbf{s}_1)u(\mathbf{s}_2)\right\rangle_{\ell=0}
    &= \frac{1}{15} f x (f-5 b) \Xi _1^{(1)}(s)  \\  
    \left\langle \delta(\mathbf{s}_1)u(\mathbf{s}_2)\right\rangle_{\ell=1}
    &= -\left(b+\frac{3f}{5}\right)\Xi_1^{(1)}(s) \nonumber \\
    &+ \frac{x^2}{20}\left(b+\frac{11f}{5}\right)\Xi_1^{(1)}(s) - \frac{2fx^2}{175}\Xi_3^{(1)}(s) \\
    \left\langle \delta(\mathbf{s}_1)u(\mathbf{s}_2)\right\rangle_{\ell=2}
    &= -\frac{1}{15} f x (f-5 b) \Xi _1^{(1)}(s)-\frac{4}{35} f^2 x \Xi _3^{(1)}(s)\\ 
    \left\langle \delta(\mathbf{s}_1)u(\mathbf{s}_2)\right\rangle_{\ell=3}
    &= \frac{f}{5}\left(2-\frac{x^2}{45}\right)\Xi_3^{(1)}(s) - \frac{x^2}{20}\left(b + \frac{11f}{5} \right)\Xi_1^{(1)}(s) \\
       \left\langle \delta(\mathbf{s}_1)u(\mathbf{s}_2)\right\rangle_{\ell=4}
    &=\frac{4}{35} f^2 x \Xi _3^{(1)}(s)\,.
\end{align}
Notice that for density-momentum cross correlation the pair of galaxy is not symmetric in $s_1 \leftrightarrow s_2$, which implies even multipoles are non zero even with a bisector LOS. This is different from the density-density correlation function, where the contribution of wide angle terms to the odd multipoles was zero for the bisector LOS, and one could therefore conclude they did not carry any additional cosmological information. For density-momentum cross even multipoles contain independent cosmological information and could in principle be exploited by the next generation of peculiar velocity surveys.

The multipoles of $\xi^{\rho\delta}$ are shown in Fig.~\ref{fig:fig_bis} for $\ell=0$ through $4$.  The $\ell=1$ and $3$ moments (left two columns) show very small deviations from the plane-parallel expression even for large opening angles.  Only above $30^\circ$ are the deviations reaching 10 per cent.  The even multipoles, which are identically zero for $\theta\to 0$, are typically a very small faction of the dipole amplitude even for large opening angles.  Once again it is only for $\theta>30^\circ$ that the corrections reach 10 per cent of the dipole result.

However we will find that the results in the endpoint parameterization will be more useful when making connection to the power spectrum and spherical harmonic results later, as well as producing very compact expressions, so we shall turn to this parameterization now.

\subsection{Endpoint}

\begin{figure*}
    \centering
    \includegraphics[width=\textwidth]{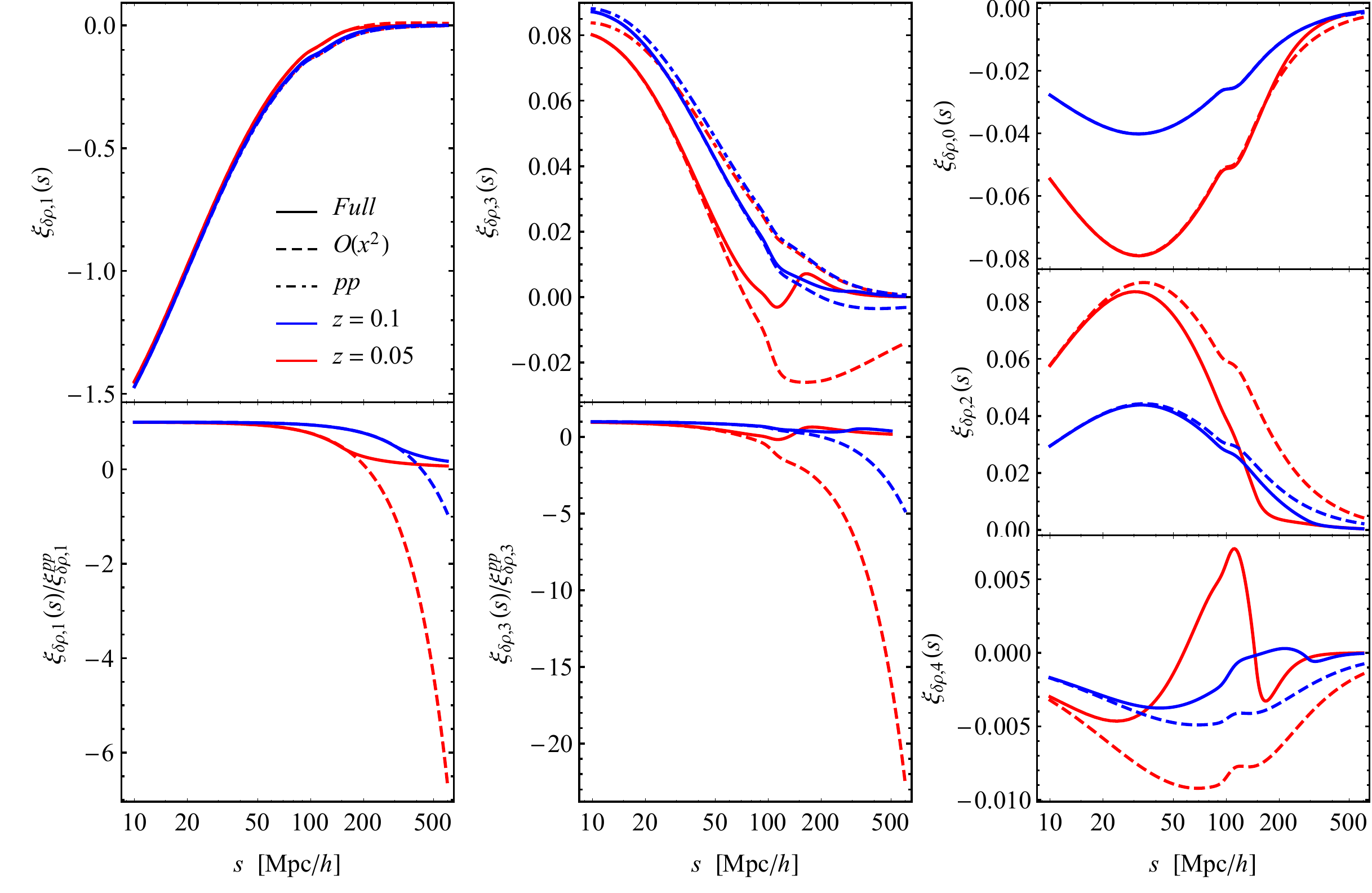}
    \caption{The multipoles of the density-momentum cross-correlation in the endpoint parameterization as a function of separation, $s$.  We have assumed $b=1$.  The left two columns show the dipole and octopole, with the upper row showing the signal and the lower row the ratio to the plane-parallel results.  The right column shows the even multipoles, which are zero in the plane-parallel limit.  The solid lines show our full expressions while the dashed lines show the expansion through $\mathcal{O}(x_1^2)$.  The small-$x_1$ expansion is most accurate for small $s$, as expected.}
\label{fig:fig_ep_dv}
\end{figure*}

\begin{figure*}
    \centering
    \includegraphics[width=\textwidth]{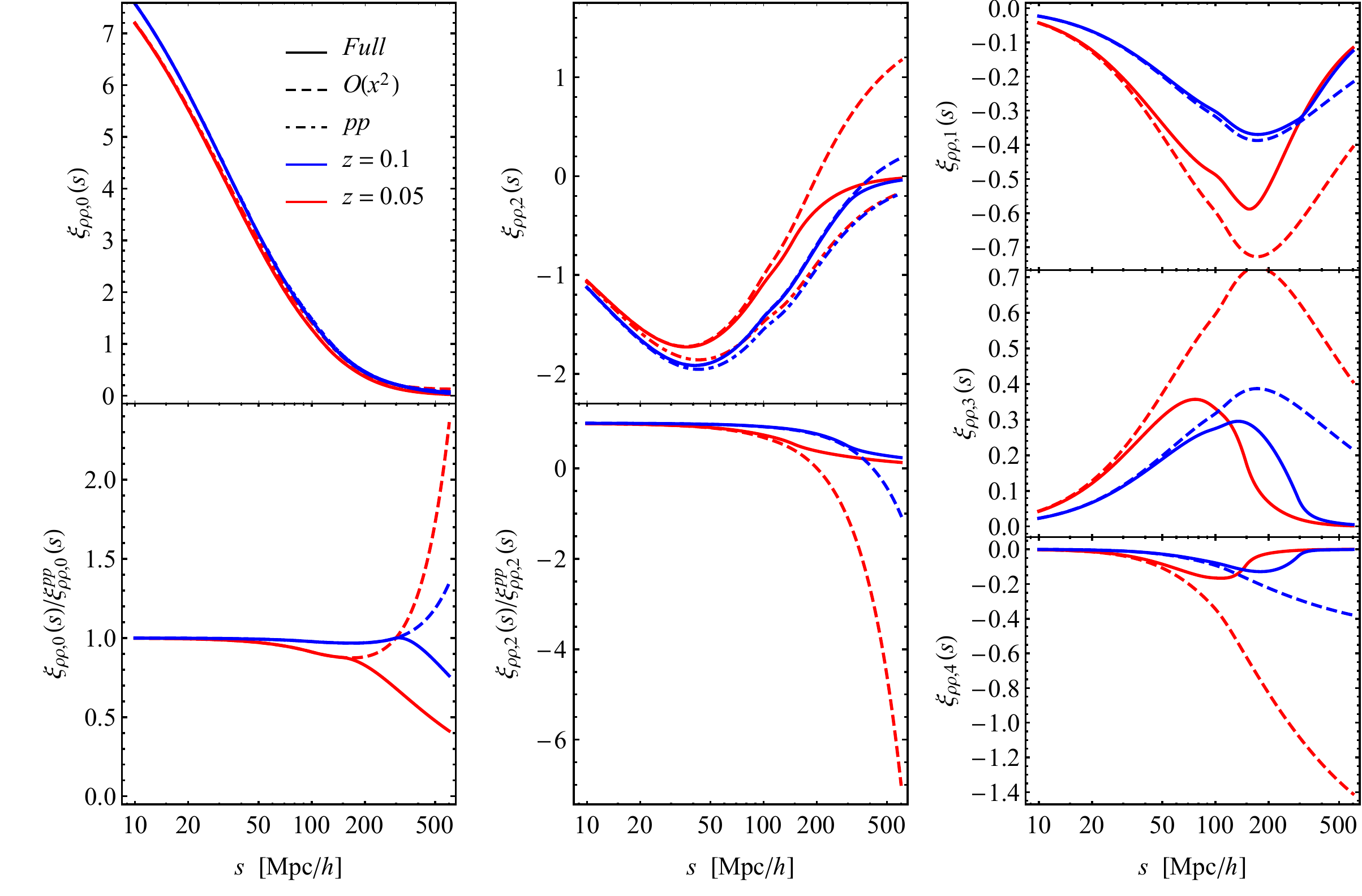}
    \caption{The multipoles of the momentum auto-correlation in the endpoint parameterization as a function of separation, $s$.  The left two columns show the monopole and quadrupole, with the upper row showing the signal and the lower row the ratio to the plane-parallel results.  The right column shows the odd multipoles, which are zero in the plane-parallel limit.  The solid lines show our full expressions while the dashed lines show the expansion through $\mathcal{O}(x_1^2)$.  The small-$x_1$ expansion is most accurate for small $s$, as expected.}
\label{fig:fig_ep_vv}
\end{figure*}

\begin{figure*}
    \centering
    \includegraphics[width=\textwidth]{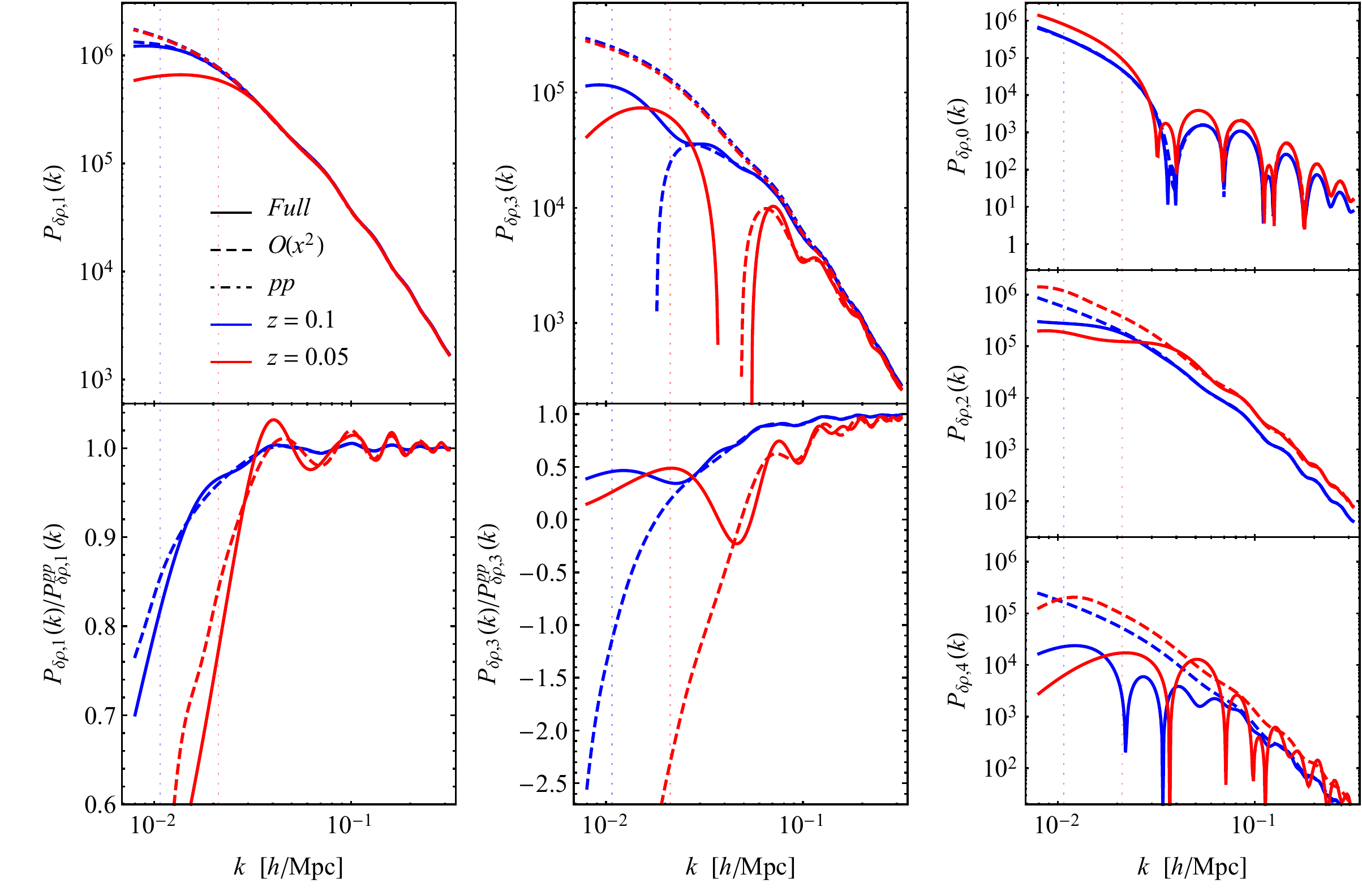}
    \caption{As for Fig.~\protect\ref{fig:fig_ep_dv} except for the power spectrum.}
\label{fig:fig_ep_dv_pk}
\end{figure*}

\begin{figure*}
    \centering
    \includegraphics[width=\textwidth]{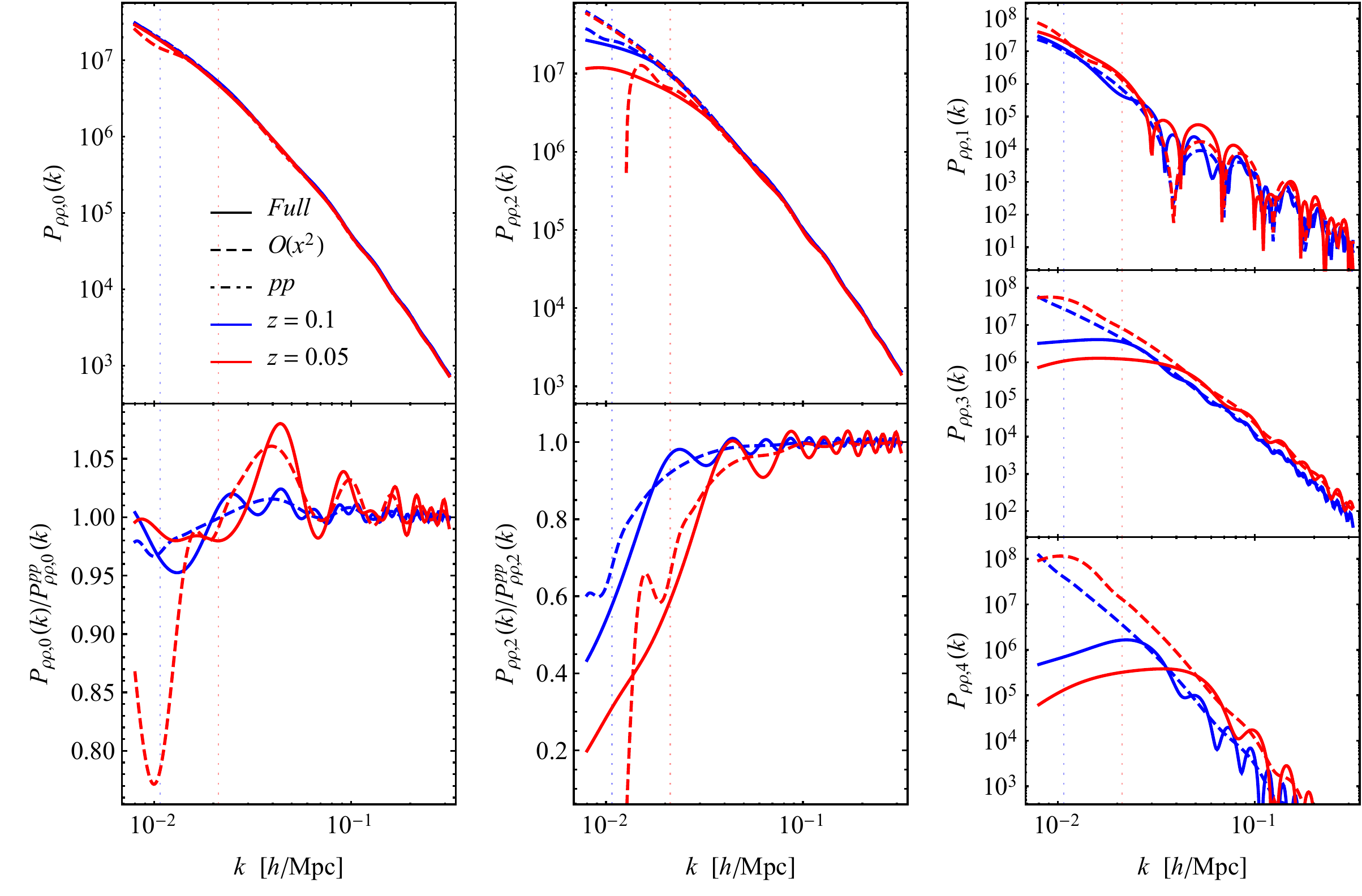}
    \caption{As for Fig.~\protect\ref{fig:fig_ep_vv} except for the power spectrum.}
\label{fig:fig_ep_vv_pk}
\end{figure*}

For the endpoint definition we take the line-of-sight to be $\mathbf{s}_1$ and our two-point functions can then be written either in terms of Eq.~\ref{eqn:deltadelta} or directly in terms of integrals of the following form
\begin{equation}
    i^L(2L+1)\int\frac{d\Omega_k}{4\pi}
    \mathcal{L}_L(\hat{k}\cdot\hat{s})
    \mathcal{L}_{\ell_1}(\hat{k}\cdot\hat{s}_1)
    \mathcal{L}_{\ell_2}(\hat{k}\cdot\hat{s}_2)
\end{equation}
We choose a coordinate system with $\hat{s}_1$ along the $z$-axis, $\hat{s}_2$ in the $x-z$ plane and $\hat{k}$ arbitrary.  The above integral can be simply evaluated by first integrating over $\phi_k$ and then over $\mu_k=\hat{k}\cdot\hat{s}_1$.
The full expression is rather compact for both density-momentum cross correlations
\begin{align}
   \left\langle \delta(\mathbf{s}_1)u(\mathbf{s}_2)\right\rangle = &  \frac{f^2 \left(\mu_1\left[5\mu_1^2+3\mu_1 x_1-3\right]-x_1\right) \Xi_3^{(1)}(s)}{5\sqrt{1+x_1^2+2\mu_1 x_1}}
   \nonumber \\ 
   & - \frac{f\Xi_1^{(1)}(s) \left(x_1\left[2f\mu_1^2+f+5\right]+(3f+5) \mu_1
   \right)}{5 \sqrt{1+x_1^2+2\mu_1 x_1}}
\end{align}
and momentum-momentum correlations
\begin{align}
     \left\langle u(\mathbf{s}_1)u(\mathbf{s}_2)\right\rangle &= \frac{f^2(1+\mu_1x_1) \Xi_0^{(2)}(s)}{3\sqrt{1+x_1^2+2\mu_1x_1}} \nonumber \\
     &+ \frac{f^2\left(1-2\mu_1x_1-3\mu_1^2\right)\Xi_2^{(2)}(s)}{3\sqrt{1+x_1^2+2\mu_1x_1}}
\end{align}
where we have defined $x_1=s/s_1$ and $\mu_1=\hat{s}_1\cdot\hat{s}$.
Our expression for $\langle uu\rangle$ agrees with the result of \citet{Gorski88}, see Appendix \ref{app:gorski88}.

To $O(x_1^2)$ the multipoles of the density-momentum correlation are
\begin{align}
    \left\langle \delta(\mathbf{s}_1)u(\mathbf{s}_2)\right\rangle_{\ell=0}
    &= f\Xi_1^{(1)}(s)\frac{2x_1}{3}\left(b+\frac{f}{5}\right) \\
    \left\langle \delta(\mathbf{s}_1)u(\mathbf{s}_2)\right\rangle_{\ell=1}
    &= f\Xi_1^{(1)}(s)\left(b\left[1-\frac{3x_1^2}{5}\right] + \frac{f}{5}\left[3-x_1^2\right]\right)
    \nonumber \\
    &+ \Xi_3^{(1)}(s)\frac{2f^2x_1^2}{35} \\
    \left\langle \delta(\mathbf{s}_1)u(\mathbf{s}_2)\right\rangle_{\ell=2}
    &= -f\Xi_1^{(1)}(s)\frac{2x_1}{3}\left(b + \frac{f}{5}\right)
    - \Xi_3^{(1)}(s)\frac{8f^2x_1}{35}
\end{align}
and so on, while the momentum auto-correlation is
\begin{align}
    \left\langle u(\mathbf{s}_1)u(\mathbf{s}_2)\right\rangle_{\ell=0}
    &= f^2\Xi_0^{(2)}(s)\left[ \frac{1}{3} - \frac{x_1^2}{9}\right] 
    + f^2\Xi_2^{(2)}(s)\left[\frac{4x_1^2}{45}\right] \\
    \left\langle u(\mathbf{s}_1)u(\mathbf{s}_2)\right\rangle_{\ell=1}
    &= -\Xi_2^{(2)}(s)\frac{2f^2x_1}{5} \\
    \left\langle u(\mathbf{s}_1)u(\mathbf{s}_2)\right\rangle_{\ell=2}
    &= f^2\Xi_0^{(2)}(s)\frac{x_1^2}{9} - f^2\Xi_2^{(2)}(s)\left[ \frac{2}{3}-\frac{16x_1^2}{63}\right]
\end{align}
etc.
Fig. \ref{fig:fig_ep_dv} shows the multipoles of the density-momentum correlation function in the endpoint parametrization assuming $s_1 = \chi(z=0.1) = 298\,h^{-1}$Mpc, blue set of curves, and $s_1 = \chi(z=0.05) = 148\,h^{-1}$Mpc, red set of curves. The continuous lines correspond to the exact result, the dashed ones to the series expansion to $\mathcal{O}(x^2)$, and the dot-dashed to the plane parallel limit which is non-zero only for the dipole and the octopole. 
We notice that for the midpoint choice of the LOS the value of $x$ is bounded, $x\le2$, but this is not the case for the endpoint LOS. 
As expected when $x\simeq 1$ the series expansion performs very poorly, and very large deviations from plane-parallel can be clearly seen for $\ell=1$, 2, 3 and $4$.
Fig. \ref{fig:fig_ep_vv} instead shows the wide angle corrections to the momentum auto power spectrum, using the same color coding as Fig.~\ref{fig:fig_ep_dv}. For all multipoles the deviations from the distant observer limit are substantial, and become larger for higher $\ell$.

\section{Power spectrum}
\label{sec:fourier}

As a result of having chosen a preferred position in the Universe (that of the observer), RSD partially break statistical homogeneity and isotropy of our 2-point functions. The only symmetry one is left with, in the absence of a window function, is rotational symmetry around the observer and azimuthal symmetry about the line of sight.  In particular the loss of translation invariance means the power spectrum is no longer diagonal and some care must be exercised in its definition \citep{ZH96,Sza98,Rei16,CasWhi18a}.
What is always well-defined is the ``local'', i.e.~line-of-sight-dependent, power spectrum \citep{Sco15,Rei16}, which in the endpoint parameterization can be written
\begin{equation}
  P(\mathbf{k},\mathbf{s}_1) \equiv \int \mathrm{d}^3 s
    \, \xi(\mathbf{s},\mathbf{s}_1)e^{-i\mathbf{k}\cdot\mathbf{s}}
\end{equation}
This can be expanded in multipoles as
\begin{equation}
  P(\mathbf{k},\mathbf{s}_1) = \sum_{L} P_L(k,s_1)
  \mathcal{L}_L\left(\hat{k}\cdot\hat{s}_1\right)\;.
\label{eqn:Pk_expansion}
\end{equation}
In observations the most commonly used estimator for the density power spectrum multipoles is a variant of the \citet{Yam06} estimator:
\begin{align}
  \hat{P}_L^{FFT}(k) &\equiv \frac{(2L+1)}{V}\int \frac{\mathrm{d}\Omega_\mathbf{k}}{4\pi}\mathrm{d}^3 s_1 \mathrm{d}^3 s_2 \notag \\
  & \times \delta(\mathbf{s}_1) \delta(\mathbf{s}_2) e^{-i\mathbf{k}\cdot \mathbf{s}} \mathcal{L}_L\left(\hat{k}\cdot\hat{s}_1\right)
\label{eqn:PkFFT}
\end{align}
which can be evaluated using FFTs \citep{Sco15,Bia15,Hand17a}.

It is common practice \citep{Park00,Howlett19} to treat the line-of-sight component of the momentum the same way as a density field, even though it is a single component of a vector (spin-1) field rather than a scalar.  In this case the ensemble average of the FFT estimator for any pair of fields can be related to the multipoles in Eq.~\ref{eqn:Pk_expansion} as \citep{CasWhi18a}:
\begin{equation}
    \left\langle\hat{P}_L^{FFT}\right\rangle = \int\frac{d^3s_1}{V} P_L^{XY}(k,s_1)
\end{equation}
where $X,Y\in \delta, u$.  
Computing the ensemble averaged multipoles of the power spectrum estimators defined above requires a Hankel transform of the expression for the correlation function
\begin{align}
\label{eq:P_ell}
    P_L(k,s_1) =  4\pi (-i)^L \int s^2\mathrm{d}{s}\, \xi_L(s,s_1)\, j_L(ks)  \,.
\end{align}
We can also then use the series expansion for the correlation function in powers of $(s/s_1)$ to define the analogous $P_L^{(n)}(k)$. 
The general expression, including angular mask and radial selection, for the measured FFT estimator is given by the convolution of the underlying theory with a window function, $W(\vec{s}_{1,2})$, and can be found in \citet{CasWhi18a,Beu19}, which we reproduce here for convenience:
\begin{align}
    \langle P_A^{\rm FFT} (k) \rangle
    &=  4\pi (-i)^A (2A+1) \sum_{\ell,\, L}\tj{\ell}{L}{A}{0}{0}{0}^2(2L+1) \notag \\\
    &   \int s^2 \mathrm{d}s\, j_A (ks) \sum_n (s)^n\, \xi_{XY,\ell}^{(n)}(s) \notag \\
	&\times \int \frac{\mathrm{d}\Omega_s}{4\pi} \int \mathrm{d}^3 s_1  (s_1)^{-n} W(\mathbf{s}_1)W(\mathbf{s}+\mathbf{s}_1)\mathcal{L}_L(\hat{s}\cdot\hat{s}_1)
\end{align}
Notice that one can use Eq.~\ref{eq:P_ell} to rewrite this in a more compact form
\begin{align}
\label{eq:PK_FFT_w}
    \langle P_A^{\rm FFT} (k) \rangle
    &=   (-i)^A (2A+1) \sum_{\ell,\, L}\tj{\ell}{L}{A}{0}{0}{0}^2(2L+1) \notag \\
    & i^L\sum_n \int \frac{k'^2\,\mathrm{d}k'}{2 \pi^2}\ P_\ell^{(n)}(k') Q_{A,\ell,L}^{(n)}(k,k')
\end{align}
where we have defined 
\begin{align}
    Q_{A,\ell,L}^{(n)}(k,k')
    &\equiv  4\pi \int \mathrm{d}s\,s^2 j_A (ks) j_\ell(k's)  (s)^n\, \notag \\
    & \int \frac{\mathrm{d} \Omega_s}{4\pi} \int \mathrm{d}^3 s_1  (s_1)^{-n} W(\mathbf{s}_1)W(\mathbf{s}+\mathbf{s}_1)\mathcal{L}_L(\hat{s}\cdot \hat{s}_1)\,
\end{align}
The above expression for the $Q_{A,\ell,L}^{(n)}(k,k')$ is rather involved, but in principle it has to be computed only once. Alternatively one could try to invert Eq.~\ref{eq:PK_FFT_w} to recover the $P_\ell(k)$, similarly to what is done in CMB analysis in the `Pseudo-$C_\ell$' method \citep{Hivon02}.

Fig. \ref{fig:fig_ep_dv_pk} shows the prediction for the multipoles of the density momentum cross power spectrum at $z=0.05$, $0.1$. For the dipole the series expansion is a good approximation to the exact expression even at very large scales, while the plane parallel limit overestimates power by tens of percent. Even multipoles of the cross power spectrum are clearly non-zero as one can see from the right column.  Lower redshifts are more affected by wide angle effects as expected. The octopule shows larger discrepancies between the $\mathcal{O}(x^2)$ and the full result, especially at $z=0.05$. 
Fig. \ref{fig:fig_ep_vv_pk} shows the five multipoles of the momentun-momentum power spectrum. As for the case of the correlation function the series expansion becomes a bad approximation when $k s_1 \simeq 1$, and one has to switch to the full result. For the monopole we find a few per cent difference between the wide angle and flat-sky formulae, which becomes 20-40 per cent for the quadrupole. The multipoles with $L=1$, 2, $4$ are non-zero because of our choice of asymmetric LOS, and could be used as a test of possible systematics in the modeling or in the data.

\section{Spherical coordinates}
\label{sec:SFB}

As RSD break the translational symmetry of the theory down into a rotational symmetry about the observer, it is natural to seek basis modes which reflect the angular-radial decomposition suggested by the theory.  One such basis is the spherical Fourier-Bessel (sFB) expansion, most frequently encountered in potential theory, or its configuration space analog, the multi-frequency angular power spectrum (MAPS; \citealt{Dat07}; see also \citealt{CasWhi18a,CasWhi18b}).  Since the original papers of \citet{Lahav94,Fisher94,HT95} several authors have studied galaxy clustering in spherical coordinates, see for instance \citet{YD2013,Pratten13,Nic14,Sha14,Liu16,Pas17,CasWhi18a,CasWhi18b,Samushia19} and references therein.
These analyses retain a clear separation between angular and radial coordinates, i.e.~redshifts. The methods have been successfully applied to data in \citet{Fisher94,Tadros99,Tay01,Pad01,Percival04,Pad07}.

Defining the multipole moments of a field on the sphere as
\begin{equation}
  X(\mathbf{s}) = \sum_{\ell m} X_{\ell m}(s) Y_{\ell m}(\hat{s})
\end{equation}
with $Y_{\ell m}$ the spherical harmonics, and using Eq.~\ref{eqn:deltas-defn} we can write the linear theory density as
\begin{equation}
    \delta_{\ell m}(s) = i^\ell \int\frac{d^3k}{2\pi^2}
    \delta(\mathbf{k})\left[ b j_\ell(ks)-f j_\ell''(ks) \right] Y_{\ell m}^{\star}(\hat{k})
\end{equation}
where the ${}^\prime$ indicate a derivative with respect to the argument (i.e.~$ks$).  Similarly the line-of-sight velocity is
\begin{equation}
    u_{\ell m}(s) = i^\ell  f \int\frac{d^3k}{2\pi^2}
    \delta(\mathbf{k}) k^{-1} j_{\ell}'(ks) Y_{\ell m}^{\star}(\hat{k})
\end{equation}
The $j_\ell'$ can also be written in terms of $j_{\ell\pm 1}$ and $j_\ell''$ in terms of $j_{\ell\pm 1}$ and $j_{\ell\pm 2}$ which makes more manifest the coupling of intrinsic angular momentum and structure generated through projection.
The two point functions of the fields are simply related to that of the configuration space multipoles 
\begin{equation}
 \xi^{XY}(\mathbf{s}_1 , \mathbf{s}_2) =
 \sum_\ell\frac{2\ell+1}{4\pi}C_\ell^{XY}(s_1,s_2) \mathcal{L}_\ell(\hat{s}_1\cdot\hat{s}_2)
\label{eq:xi_from_Cl}
\end{equation}
where $X,Y\in \delta, u$.
This makes clear that this formalism describes the triangle of Fig.~\ref{fig:triangle} in terms of the two side lengths ($s_1$ and $s_2$) and the enclosed angle and then expands the angular dependence in Legendre polynomials.  The coefficients, $C_\ell$, are the MAPS and their one dimensional Hankel transform along $s_1$ and $s_2$ the angular power spectra $C_\ell(k_1,k_2)$.
A straightforward calculation shows that in linear theory
\begin{align}
 C_\ell^{\delta\delta}(s_1, s_2) &= \frac{2}{\pi} \int k^2\mathrm{d}\,k\, P(k)
 \left[b j_\ell(ks_1)-fj_\ell''(ks_1)\right] \nonumber \\
 &\times \left[b j_\ell(ks_2)-fj_\ell''(ks_2)\right]   .
\label{eq:Csell}
\end{align}
The auto-spectrum of the velocity becomes
\begin{equation}
 C_\ell^{uu}(s_1, s_2) = \frac{2f^2}{\pi} \int\mathrm{d}\,k\, P(k)j_\ell'(ks_1)j_\ell'(ks_2)
\end{equation}
while the density-velocity cross spectrum is
\begin{align}
 C_\ell^{\delta u}(s_1, s_2) &= \frac{2f}{\pi} \int k\mathrm{d}\,k\, P(k) \nonumber \\
 &\times \left[b j_\ell(ks_1)-fj_\ell''(ks_1)\right]j_\ell'(ks_2)
\end{align}
These spectra have power over a wide range of $\ell$.  However at large scales most of the intrinsic power is confined in a few multipoles, $L$, and thus the complex structure of the MAPS or angular power spectra predominantly results from projection effects \citep{CasWhi18b}, similar to what happens with the CMB where only a few multipoles are relevant at recombination and the rich structure we observe today is due to LOS projection \citep{Hu97,Dod03}.

To see this let us relate $C_\ell$ to the power spectrum with respect to the end-point definition.  We can write \citep{CasWhi18b}
\begin{equation}
  C_\ell^{XY}(s_1,s_2) = \sum_{L\lambda} F^{\ell}_{L\lambda}\int\frac{k^2\mathrm{d}k}{2\pi^2}\, P_L^{XY}(k,s_1) j_{\lambda}(ks_1) j_\ell(ks_2)
\label{eqn:Clxy}
\end{equation}
where
\begin{align}
  F^\ell_{L\lambda} &= 4\pi(2\lambda+1)i^{\lambda-\ell}\tj{\ell}{L}{\lambda}{0}{0}{0}^2
  \quad .
\end{align}
On sufficiently large scales the `intrinsic' power spectrum, $P_L(k,s_1)$, is only non-negligible for a small number of $L$.  The triangle condition of the $3j$-coefficients makes the sum over $\lambda$ finite, as $|\lambda-\ell|\le L$.  The power in multiple $\ell$ thus arises from geometric projection.  Identities which can help in the evaluation of Eq.~\ref{eqn:Clxy} can be found in \citet{CasWhi18a,CasWhi18b}.

As well as providing physical insight, Eq.~\ref{eqn:Clxy} also solves one of the major issues of spherical analysis: the estimate of the covariance matrix.  Since a typical analysis would have hundreds of $\ell$ modes and tens of $s$ or $k$ bins, each of which can be highly covariant, the accuracy of the covariance matrix presents a challenge \citep[e.g.][]{Percival04}.
Eq.~\ref{eqn:Clxy} provides a solution, as the $C_\ell$ and the $P_L$ are linearly related to each other by a matrix that can be `inverted' to find an optimal data compression.  Additionally, 
Eq.~\ref{eqn:Clxy} provides an elegant and unbiased way to remove systematics in the plane of the sky, that by definition affect only the low-$k_\parallel$ modes.

\section{Zeldovich approximation}
\label{sec:zeldovich}

\begin{figure}
    \centering
    \includegraphics[width=.425\textwidth]{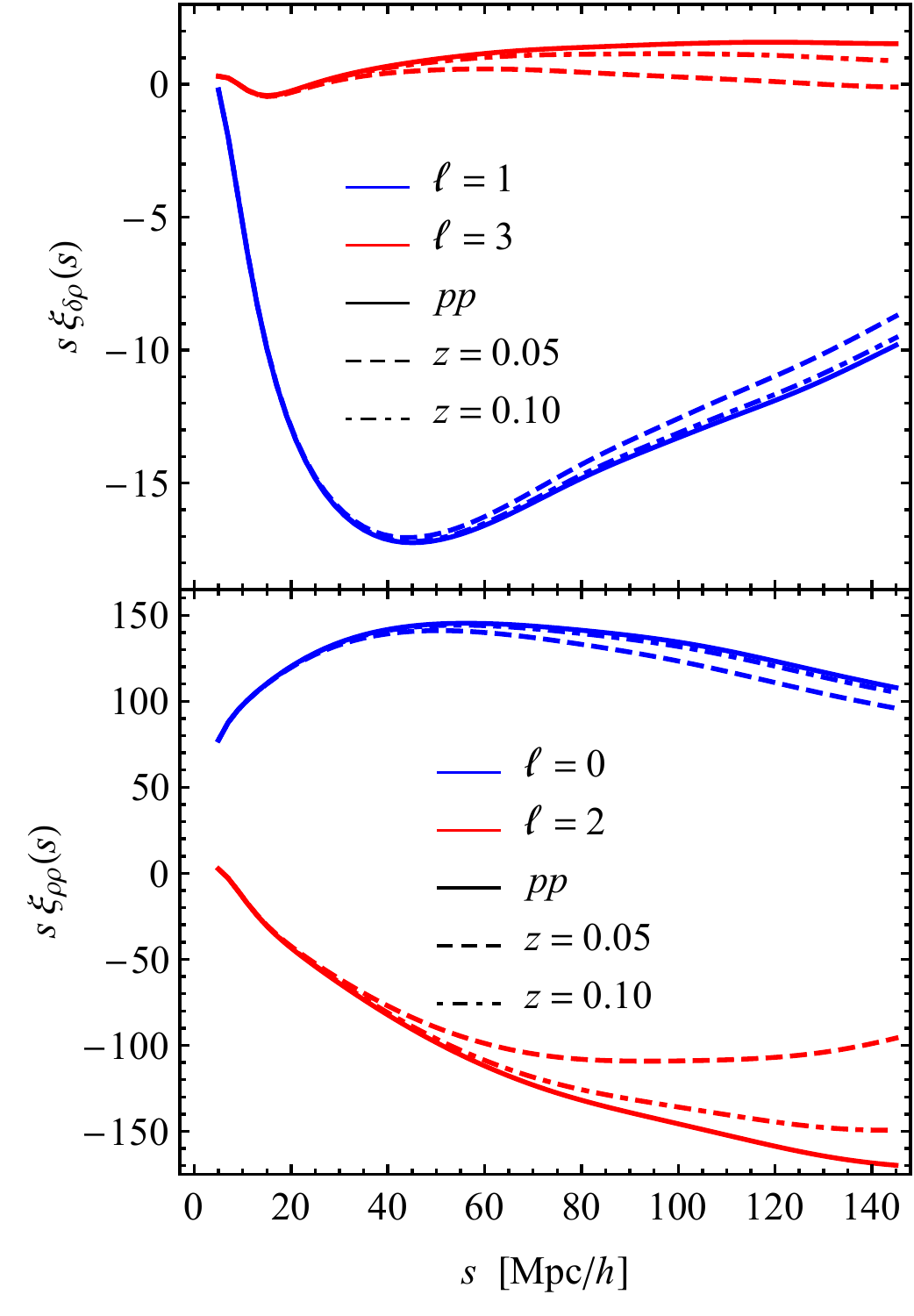}
    \caption{Top panel: Multipoles of the momentum-density cross-correlation function in the endpoint parameterization, computed using the Zeldovich approximation.  We show results for $z = 0.05,\, 0.10$ and the plane parallel case.  (Bottom) The line-of-sight momentum density auto-correlation in the Zeldovich approximation for the same configurations.
    }
\label{fig:zeldovich}
\end{figure}

It is straightforward to include the dynamical wide-angle effects in lowest order Lagrangian perturbation theory, i.e.~the Zeldovich approximation \citep{Zel70}.  Such an approach can include a more complex bias model as well as properly resumming the large-scale displacements that are not well captured by linear perturbation theory \citep[see the discussion in][]{CasWhi18b,Taruya}.

The momentum-density and momentum-momentum correlation functions are computed from generating functions in the usual way \citep[see e.g.][for examples]{Wang14}, and expressions in the plane-parallel limit can be found in \citet{Hand17b}.  These can be straightforwardly modified to include wide-angle effects following \citet{CasWhi18b} with the final integrals performed numerically without relying on the series expansion in $x$.  The large-scale limit of the Zeldovich approximation agrees with our linear theory expressions, as shown in Appendix \ref{app:Zel2Lin}.

The upper panel in Fig.~\ref{fig:zeldovich} shows the $\ell=1$ and 3 multipoles of the momentum-density cross-correlation function, computed within the endpoint parameterization, for pairs with $s_1=150$, 300 and for the plane parallel case.  For the closest galaxies the wide angle effects can reach 10 per cent at large pair separations.
The lower panel shows the $\ell=0$, $2$ multipoles of the momentum auto-correlation function for the same configurations. For the quadrupole the wide angle effects can be as large as the plane parallel piece.
At such low redshift the shape of the ZA multipoles is different from linear theory even on large scales. Nonetheless the relative importance of wide angle effects is qualitatively similar to the linear theory calculations presented in the previous sections. 
In Fig.~\ref{fig:zeldovich} we have not included the ``selection function terms'' of Appendix \ref{app:alpha}.  For smoothly varying $\bar{n}(\mathbf{s})$ these terms can be added using linear theory, and behave precisely as described in the previous sections.

\section{Conclusions}
\label{sec:conclusions}

In this paper we investigated the impact of wide angle redshift space distortions on two point statistics of the density and momentum fields within linear theory and the Zeldovich approximation.

We presented results for the linear theory correlation function (Figs.~\ref{fig:fig_bis}, \ref{fig:fig_ep_dv} and \ref{fig:fig_ep_vv}) and power spectrum (Figs.~\ref{fig:fig_ep_dv_pk} and \ref{fig:fig_ep_vv_pk}) of the cross correlation between density and momenta and for the momentum auto-correlation. In both cases wide angle effects are significant at low redshift and could become important in the analysis of upcoming surveys like Taipan.
For the cross-correlation between density and velocity we point out that the even multipoles, which would be zero in the plane parallel limit, carry new cosmological information as they are non-zero regardless of the choice of LOS. 

We were able to compute the full wide angle correlation functions and power spectra, without relying on a series expansion in small angles (\S\ref{sec:fourier}).  This turned out to be important for high multipoles, where the asymptotic nature of the expansion makes the prediction  very inaccurate at large scales. 

The manner in which redshift-space distortions behave makes a discussion in spherical coordinates particularly appealing.  We described the formalism for peculiar velocity statistics in the Fourier-Bessel basis in \S\ref{sec:SFB}, explicitly discussing the connection with the configuration and Fourier space pictures and the generation of large $\ell$ power through projection and aliasing.  The link to the Fourier description, in which the power is localized in a small number of multipoles, provides a route for efficiently compressing the information and regularizing the covariance matrix in the FB basis.

Finally we computed the dynamical part of wide angle effects in the Zeldovich approximation (\S\ref{sec:zeldovich}; Fig.~\ref{fig:zeldovich}), finding very similar conclusions with respect to the linear theory calculation which we show it approaches for sufficiently large scales (Appendix \ref{app:Zel2Lin}).  For slowly varying mean density the terms coming from the change in volume between real and redshift space, known as $\alpha$-terms in the literature, can be included using linear theory.

\vspace{0.2in}
The authors thank Enzo Branchini for discussions and comments on the draft. M.W.~is supported by the U.S.~Department of Energy and by NSF grant number 1713791.
This work made extensive use of the NASA Astrophysics Data System and of the {\tt astro-ph} preprint archive at {\tt arXiv.org}. 

\appendix

\section{Mean density terms}
\label{app:alpha}

Outside of the plane-parallel approximation a variation of the volume element and mean density with distance lead to corrections which scale as $1/s$ \cite{Kai87,H98}.  \citet{Sza98} refer to these terms as ``$\alpha$ terms'' and we shall follow their lead.  The $\alpha$ terms are collected in this Appendix.

\begin{figure}
    \centering
    \includegraphics[width=0.235\textwidth]{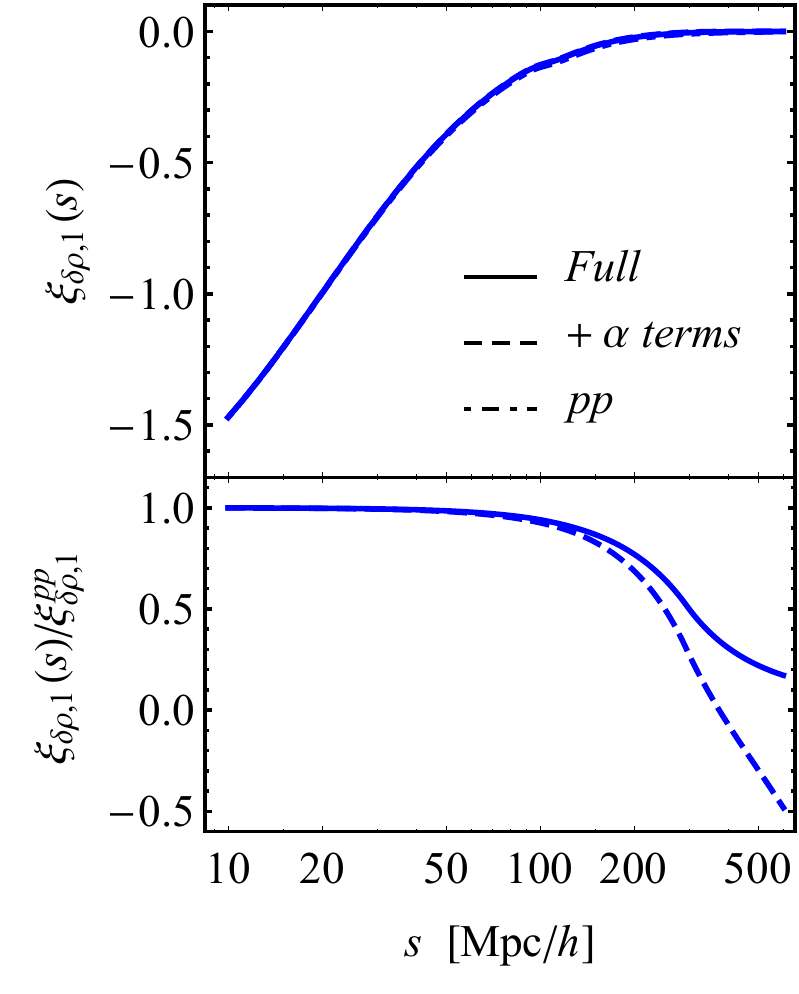}
    \includegraphics[width=0.235\textwidth]{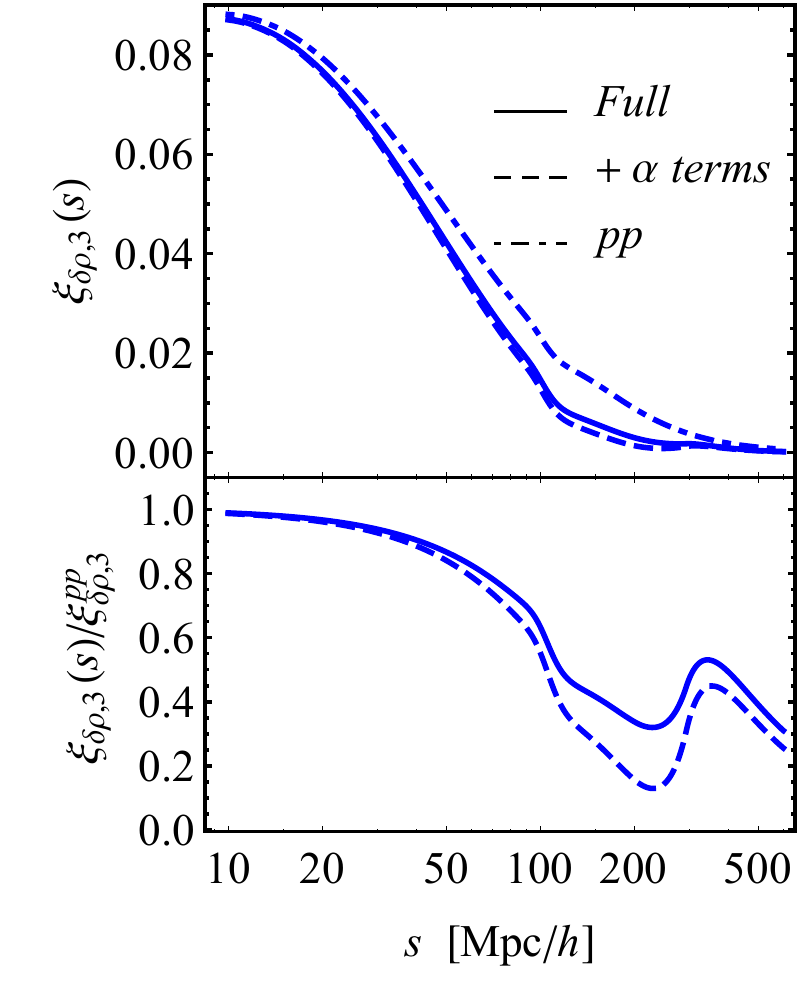}
    \caption{The density momentum correlation function at $z=0.1$ including the effect of a constant, in redshift, selection function $\alpha(s)=2/s$.}
    \label{fig:alpha}
\end{figure}

We use the common definition that $\alpha(\mathbf{r})$ is the logarithmic derivative of the galaxy selection function
\begin{equation}
  \alpha(r)\equiv \frac{d\ln r^2\bar{n}(r)}{d\ln r}
\end{equation}
which we will assume varies slowly with $r$.  The redshift-space density field then gets an additional contribution $-if\alpha\delta_1^1$ while the velocity field is unchanged (in linear theory).  The contributions to the density auto-correlation function can be found in Eqs.~(18-21) of \citet{Sza98}.  The momentum-density cross-correlation picks up an additional term
\begin{equation}
    \left\langle \delta(\mathbf{s}_1) u(\mathbf{s}_2)\right\rangle \ni  f^2\alpha(s_1)\langle\delta_1^1(\vb{s}_1)\delta_1^1(\vb{s}_2)\rangle = - \alpha(s_1) \langle u(\vb{s}_1)  u (\vb{s}_2)\rangle
\end{equation}
and it is therefore proportional to the $\langle uu\rangle$ term.  The $\langle uu\rangle$ term itself is unchanged. 
In \citet{CasWhi18a} it was found that for a constant selection function, \ie $\alpha(s) = 2/s$, the new terms partially canceled the dynamical wide angle effects in the density-density correlation function and power spectra \citep[see also][]{Taruya}. It is therefore interesting to see what happens to the density-momentum correlators discussed in this paper. Fig \ref{fig:alpha} shows the odd multipoles of the density-momentum correlation function at $z=0.1$, including the $\alpha$ terms (dashed lines). They are indeed comparable to the wide angle effects and for the dipole they eventually become the largest correction to the plane parallel limit. 
We notice that the for a constant selection function, the extra terms are sometimes referred to as Doppler terms \citep[e.g.][]{Bonvin11,Raccanelli2016}.

\section{Comparison with Gorski (1988)}
\label{app:gorski88}

In linear theory the real-space and redshift-space $\langle uu\rangle$ are the same, and the real-space result was first published by \citet{Gorski88}.  In the  notation of \citet{Gorski88}, his Eq.~(1), we have
\begin{equation}
    \left\langle u(\mathbf{s}_1)u(\mathbf{s}_2)\right\rangle = \Psi_\perp(s)(\hat{s}_1\cdot\hat{s}_2) + \left[\Psi_\parallel(s) - \Psi_\perp(s) \right](\hat{s}_1\cdot\hat{s})(\hat{s}_2\cdot\hat{s})
\end{equation}
with
\begin{equation}
    \Psi_\perp(s)     =  \frac{f^2}{3}\left[\Xi_0^{(2)} +  \Xi_2^{(2)}\right]  , 
    \Psi_\parallel(s) =  \frac{f^2}{3}\left[\Xi_0^{(2)} - 2\Xi_2^{(2)}\right]
\end{equation}
upon substitution of $3j_1(x)/x=j_0(x)+j_2(x)$ in his expressions.  Thus,
\begin{equation}
    \left\langle u(\mathbf{s}_1)u(\mathbf{s}_2)\right\rangle = f^2\frac{\cos\theta}{3}\left[\Xi_0^{(2)}  + \Xi_2^{(2)} \right]- \frac{f^2}{3} \Xi_2^{(2)}(\hat{s}_1\cdot\hat{s})(\hat{s}_2\cdot\hat{s})
\end{equation}
This is most easily evaluated in the enpoint parameterization.  Taking $\mathbf{s}_1=s_1(0,0,1)$, $\mathbf{s}=x_1s_1(\sqrt{1-\mu_1^2},0,\mu_1)$ and $\mathbf{s}_2=\mathbf{s}_1+\mathbf{s}$ (\citealt{Gorski88} defines $\mathbf{s}=\mathbf{s}_2-\mathbf{s}_1$) so that $s_2=s_1\sqrt{1+x_1^2+2\mu_1 x_1}$ we find
\begin{align}
    \left\langle u(\mathbf{s}_1)u(\mathbf{s}_2)\right\rangle &=  \frac{f^2(1+\mu_1 x_1)}{3 \sqrt{1+x_1^2+2\mu_1 x_1}}\Xi_0^{(2)}(s) \nonumber \\
    &+ \frac{f^2 \left(1-2\mu_1 x_1 - 3 \mu_1^2\right) }{3 \sqrt{1+x_1^2+2\mu_1 x_1}} \Xi_2^{(2)}(s) \,,
\end{align}
with $x_1=s/s_1$ and $\mu_1=\hat{s}\cdot\hat{s}_1$.  This matches the result in the main text.

\section{The exact expression for the density-density correlation function}
\label{app:density}

\begin{figure*}
    \centering
    \includegraphics[width=\textwidth]{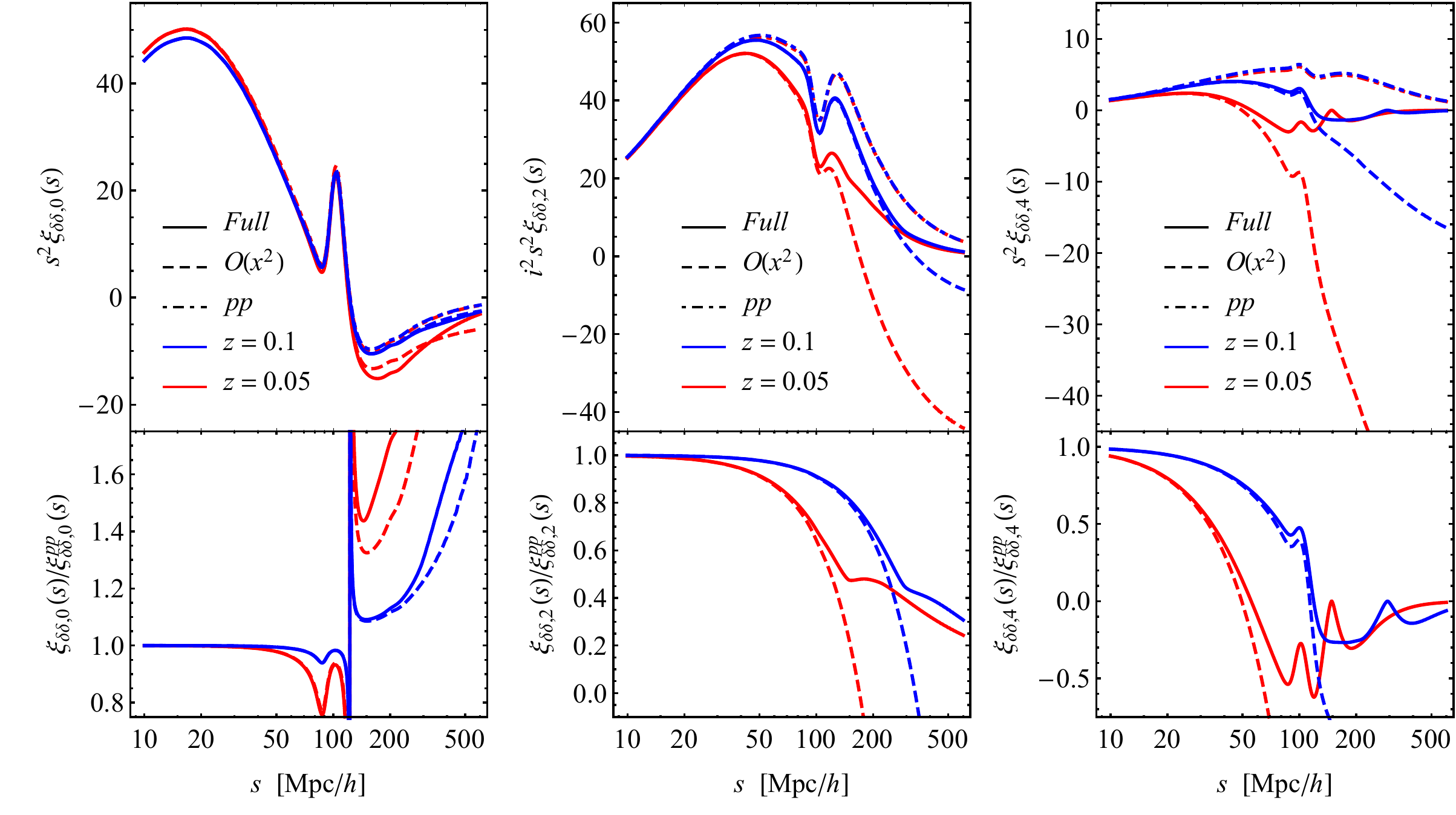}
    \caption{The even multipoles of the density-density correlation function for the same configurations of Fig \ref{fig:fig_ep_dv}.}
    \label{fig:xi_dd}
\end{figure*}

\begin{figure*}
    \centering
    \includegraphics[width=\textwidth]{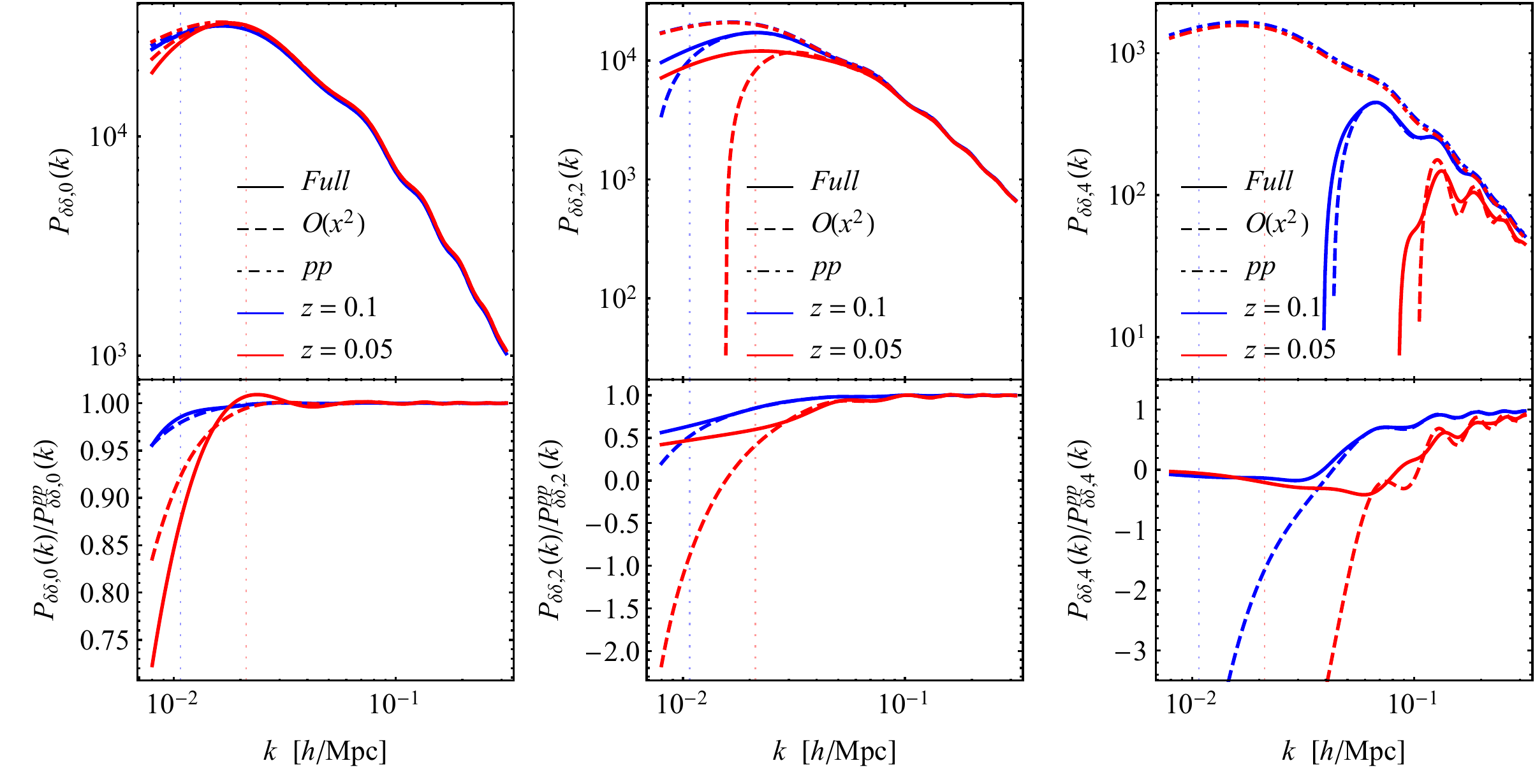}
    \caption{The even multipoles of the density-density power spectrum for the same configurations of Fig \ref{fig:fig_ep_dv_pk}.}
    \label{fig:pk_dd}
\end{figure*}
In the main text we have shown how the series expansion in $x_1=s/s_1$ becomes a poor approximation at low redshift for momentum statistics. 
In this Appendix we show the same is true for the density-density correlation function.

In Fig.~\ref{fig:xi_dd} we plot, for the endpoint choice of LOS, the even multipoles of the density auto correlation function for the same configurations in Fig.~\ref{fig:fig_ep_dv}. For the monopole, on scales smaller than the BAO scale, wide angle effects are a few per cent at $z=0.1$ and a few tens of per cent at $z=0.05$, but the series expansions performs well. For higher multipoles wide angle effects become larger and the truncation to second order in $x$ provides a very poor description of the multipoles. 
The same conclusions apply to the power spectra, shown in Fig \ref{fig:pk_dd}. These results extend the work of \citet{CasWhi18a} who provided expressions for the multipoles of the density auto-power spectrum to $\mathcal{O}(k s_1)^{-2}$.

In a joint analysis of momentum and velocity statistics at $z \lesssim 0.1$ all expressions should therefore be computed exactly.

\section{Linear theory as a limit of the Zeldovich Approximation}
\label{app:Zel2Lin}

For very large scales we expect the Zeldovich approximation to approach the linear theory results.  The relevant expressions for the density auto-correlations can be found in \citet{CasWhi18b}, and in this appendix we provide the limits for the density-momentum cross-spectrum and the velocity auto-spectrum for the case of unbiased tracers (the bias terms proceed similarly).

The two-point statistics in the Zeldovich approximation can be written as an integral of a `source' term over a Gaussian piece, e.g.
\begin{equation}
    \left\langle \rho(\vb{s}_1)\delta (\vb{s}_2)\right\rangle = \int d^3q\int\frac{d^3k}{(2\pi)^3}e^{i\mathbf{k}\cdot(\mathbf{s}-\mathbf{q})}
    \ G(\mathbf{k},\mathbf{q}) \mathcal{S}_{\rho\delta}(\mathbf{k},\mathbf{q})
\end{equation}
where the Gaussian $G(\mathbf{k},\mathbf{q})\propto \exp\left[-\frac{1}{2}k_iA_{ij}k_j\right]$ with $A(\mathbf{q})=\left\langle\left(R_1\Psi_1-R_2\Psi_2\right)^2\right\rangle$ and to lowest order the source term is
\begin{equation}
    \mathcal{S}_{\rho\delta}(\mathbf{k},\mathbf{q}) = ik\cdot\left\langle\left(R_1\Psi_1-R_2\Psi_2\right) \dot{\Psi}_1\right\rangle\cdot\hat{s}_1
\end{equation}
for the density-momentum cross-correlation and
\begin{equation}
    \mathcal{S}_{\rho\rho}(\mathbf{k},\mathbf{q}) = \hat{s}_1\left\langle\dot{\Psi}_1\dot{\Psi}_2\right\rangle\hat{s}_2
\end{equation}
for the momentum auto-correlation.  For very large scales we can expand $G(\vb{k},\vb{q}$) and set it to 1 at lowest order. The remaining integrals give the linear theory expressions of the main text.  The case of the velocity auto-correlation is straightforward to check, since $\langle\dot{\Psi}\dot{\Psi}\rangle$ is identical to $\langle uu\rangle$ in linear theory as $\Psi(\mathbf{k})=(i\mathbf{k}/k^2)\ \delta(\mathbf{k})$.  The density-momentum expression takes only a few more steps.  To begin, note that in the plane-parallel limit $\hat{s}_1\approx\hat{s}_2\approx\hat{z}$ so for example
\begin{align}
  \left\langle \rho(\vb{s}_1)\delta (\vb{s}_2)\right\rangle
  &\approx \int\frac{d^3k}{(2\pi)^3}e^{i\mathbf{k}\cdot\mathbf{s}} \ ifk\cdot\mathrm{FT}\left[R\Psi_1\Psi_2\right]\cdot\hat{z} \\
  &= \int\frac{d^3k}{(2\pi)^3}e^{i\mathbf{k}\cdot\mathbf{s}}\ \frac{if\mu}{k}(1+f\mu^2)P(k)
\end{align}
since the Fourier transform of $\left\langle\Psi_1\Psi_2\right\rangle$ is $-(k_ik_j/k^4)P(k)$ and $R_{ij}=\delta_{ij}+f\hat{z}_i\hat{z}_j$.  This matches Eq.~(\ref{eqn:rhodelta-pp}) for $b=1$.  In going beyond the plane-parallel limit one needs to keep track of the factors of $\hat{k}\cdot\hat{s}_1$ and $\hat{k}\cdot\hat{s}_2$, but it can quickly be verified that they are the same as in the main text with the velocity contributing $\hat{k}\cdot\hat{s}/k$ and the density $\left(1+[\hat{k}\cdot\hat{s}]^2\right)$ times $\delta(\mathbf{k})$.

\bibliography{}

\begin{thebibliography}{}
\makeatletter
\relax
\def\mn@urlcharsother{\let\do\@makeother \do\$\do\&\do\#\do\^\do\_\do\%\do\~}
\def\mn@doi{\begingroup\mn@urlcharsother \@ifnextchar [ {\mn@doi@}
  {\mn@doi@[]}}
\def\mn@doi@[#1]#2{\def\@tempa{#1}\ifx\@tempa\@empty \href
  {http://dx.doi.org/#2} {doi:#2}\else \href {http://dx.doi.org/#2} {#1}\fi
  \endgroup}
\def\mn@eprint#1#2{\mn@eprint@#1:#2::\@nil}
\def\mn@eprint@arXiv#1{\href {http://arxiv.org/abs/#1} {{\tt arXiv:#1}}}
\def\mn@eprint@dblp#1{\href {http://dblp.uni-trier.de/rec/bibtex/#1.xml}
  {dblp:#1}}
\def\mn@eprint@#1:#2:#3:#4\@nil{\def\@tempa {#1}\def\@tempb {#2}\def\@tempc
  {#3}\ifx \@tempc \@empty \let \@tempc \@tempb \let \@tempb \@tempa \fi \ifx
  \@tempb \@empty \def\@tempb {arXiv}\fi \@ifundefined
  {mn@eprint@\@tempb}{\@tempb:\@tempc}{\expandafter \expandafter \csname
  mn@eprint@\@tempb\endcsname \expandafter{\@tempc}}}

\bibitem[\protect\citeauthoryear{{Adams} \& {Blake}}{{Adams} \&
  {Blake}}{2017}]{Adams17}
{Adams} C.,  {Blake} C.,  2017, \mn@doi [\mnras] {10.1093/mnras/stx1529}, \href
  {https://ui.adsabs.harvard.edu/abs/2017MNRAS.471..839A} {471, 839}

\bibitem[\protect\citeauthoryear{{Amendola} et~al.,}{{Amendola}
  et~al.}{2018}]{Amendola18}
{Amendola} L.,  et~al., 2018, \mn@doi [Living Reviews in Relativity]
  {10.1007/s41114-017-0010-3}, \href
  {https://ui.adsabs.harvard.edu/abs/2018LRR....21....2A} {21, 2}

\bibitem[\protect\citeauthoryear{{Beutler}, {Castorina}  \& {Zhang}}{{Beutler}
  et~al.}{2019}]{Beu19}
{Beutler} F.,  {Castorina} E.,   {Zhang} P.,  2019, \mn@doi [\jcap]
  {10.1088/1475-7516/2019/03/040}, \href
  {https://ui.adsabs.harvard.edu/abs/2019JCAP...03..040B} {2019, 040}

\bibitem[\protect\citeauthoryear{{Bianchi}, {Gil-Mar{\'{\i}}n}, {Ruggeri}  \&
  {Percival}}{{Bianchi} et~al.}{2015}]{Bia15}
{Bianchi} D.,  {Gil-Mar{\'{\i}}n} H.,  {Ruggeri} R.,   {Percival} W.~J.,  2015,
  \mn@doi [\mnras] {10.1093/mnrasl/slv090}, \href
  {http://adsabs.harvard.edu/abs/2015MNRAS.453L..11B} {453, L11}

\bibitem[\protect\citeauthoryear{{Bonvin} \& {Durrer}}{{Bonvin} \&
  {Durrer}}{2011}]{Bonvin11}
{Bonvin} C.,  {Durrer} R.,  2011, \mn@doi [\prd] {10.1103/PhysRevD.84.063505},
  \href {http://adsabs.harvard.edu/abs/2011PhRvD..84f3505B} {84, 063505}

\bibitem[\protect\citeauthoryear{{Carrick}, {Turnbull}, {Lavaux}  \&
  {Hudson}}{{Carrick} et~al.}{2015}]{Carrick15}
{Carrick} J.,  {Turnbull} S.~J.,  {Lavaux} G.,   {Hudson} M.~J.,  2015, \mn@doi
  [\mnras] {10.1093/mnras/stv547}, \href
  {https://ui.adsabs.harvard.edu/abs/2015MNRAS.450..317C} {450, 317}

\bibitem[\protect\citeauthoryear{{Castorina} \& {White}}{{Castorina} \&
  {White}}{2018a}]{CasWhi18a}
{Castorina} E.,  {White} M.,  2018a, \mn@doi [\mnras] {10.1093/mnras/sty410},
  \href {http://adsabs.harvard.edu/abs/2018MNRAS.476.4403C} {476, 4403}

\bibitem[\protect\citeauthoryear{{Castorina} \& {White}}{{Castorina} \&
  {White}}{2018b}]{CasWhi18b}
{Castorina} E.,  {White} M.,  2018b, \mn@doi [\mnras] {10.1093/mnras/sty1437},
  \href {https://ui.adsabs.harvard.edu/abs/2018MNRAS.479..741C} {479, 741}

\bibitem[\protect\citeauthoryear{{Datta}, {Choudhury}  \& {Bharadwaj}}{{Datta}
  et~al.}{2007}]{Dat07}
{Datta} K.~K.,  {Choudhury} T.~R.,   {Bharadwaj} S.,  2007, \mn@doi [\mnras]
  {10.1111/j.1365-2966.2007.11747.x}, \href
  {http://adsabs.harvard.edu/abs/2007MNRAS.378..119D} {378, 119}

\bibitem[\protect\citeauthoryear{{Dodelson}}{{Dodelson}}{2003}]{Dod03}
{Dodelson} S.,  2003, {Modern cosmology}

\bibitem[\protect\citeauthoryear{{Dupuy}, {Courtois}  \& {Kubik}}{{Dupuy}
  et~al.}{2019}]{Dupuy19}
{Dupuy} A.,  {Courtois} H.~M.,   {Kubik} B.,  2019, \mn@doi [\mnras]
  {10.1093/mnras/stz901}, \href
  {https://ui.adsabs.harvard.edu/abs/2019MNRAS.486..440D} {486, 440}

\bibitem[\protect\citeauthoryear{{Feix}, {Branchini}  \& {Nusser}}{{Feix}
  et~al.}{2017}]{Feix17}
{Feix} M.,  {Branchini} E.,   {Nusser} A.,  2017, \mn@doi [\mnras]
  {10.1093/mnras/stx566}, \href
  {https://ui.adsabs.harvard.edu/abs/2017MNRAS.468.1420F} {468, 1420}

\bibitem[\protect\citeauthoryear{{Fisher}, {Scharf}  \& {Lahav}}{{Fisher}
  et~al.}{1994}]{Fisher94}
{Fisher} K.~B.,  {Scharf} C.~A.,   {Lahav} O.,  1994, \mn@doi [\mnras]
  {10.1093/mnras/266.1.219}, \href
  {http://adsabs.harvard.edu/abs/1994MNRAS.266..219F} {266, 219}

\bibitem[\protect\citeauthoryear{{Gorski}}{{Gorski}}{1988}]{Gorski88}
{Gorski} K.,  1988, \mn@doi [\apjl] {10.1086/185255}, \href
  {https://ui.adsabs.harvard.edu/abs/1988ApJ...332L...7G} {332, L7}

\bibitem[\protect\citeauthoryear{{Hamilton}}{{Hamilton}}{1998}]{H98}
{Hamilton} A.~J.~S.,  1998, in {Hamilton} D.,  ed.,  Astrophysics and Space
  Science Library Vol. 231, The Evolving Universe. p.~185 (\mn@eprint {}
  {astro-ph/9708102}), \mn@doi{10.1007/978-94-011-4960-0_17}

\bibitem[\protect\citeauthoryear{{Hand}, {Li}, {Slepian}  \& {Seljak}}{{Hand}
  et~al.}{2017a}]{Hand17a}
{Hand} N.,  {Li} Y.,  {Slepian} Z.,   {Seljak} U.,  2017a, \mn@doi [\jcap]
  {10.1088/1475-7516/2017/07/002}, \href
  {http://adsabs.harvard.edu/abs/2017JCAP...07..002H} {7, 002}

\bibitem[\protect\citeauthoryear{{Hand}, {Seljak}, {Beutler}  \& {Vlah}}{{Hand}
  et~al.}{2017b}]{Hand17b}
{Hand} N.,  {Seljak} U.,  {Beutler} F.,   {Vlah} Z.,  2017b, \mn@doi [\jcap]
  {10.1088/1475-7516/2017/10/009}, \href
  {https://ui.adsabs.harvard.edu/abs/2017JCAP...10..009H} {2017, 009}

\bibitem[\protect\citeauthoryear{{Heavens} \& {Taylor}}{{Heavens} \&
  {Taylor}}{1995}]{HT95}
{Heavens} A.~F.,  {Taylor} A.~N.,  1995, \mn@doi [\mnras]
  {10.1093/mnras/275.2.483}, \href
  {http://adsabs.harvard.edu/abs/1995MNRAS.275..483H} {275, 483}

\bibitem[\protect\citeauthoryear{{Hivon}, {G{\'o}rski}, {Netterfield}, {Crill},
  {Prunet}  \& {Hansen}}{{Hivon} et~al.}{2002}]{Hivon02}
{Hivon} E.,  {G{\'o}rski} K.~M.,  {Netterfield} C.~B.,  {Crill} B.~P.,
  {Prunet} S.,   {Hansen} F.,  2002, \mn@doi [The Astrophysical Journal]
  {10.1086/338126}, \href
  {https://ui.adsabs.harvard.edu/abs/2002ApJ...567....2H} {567, 2}

\bibitem[\protect\citeauthoryear{{Howlett}}{{Howlett}}{2019}]{Howlett19}
{Howlett} C.,  2019, \mn@doi [\mnras] {10.1093/mnras/stz1403}, \href
  {https://ui.adsabs.harvard.edu/abs/2019MNRAS.tmp.1367H} {p.~1367}

\bibitem[\protect\citeauthoryear{{Howlett}, {Staveley-Smith}  \&
  {Blake}}{{Howlett} et~al.}{2017a}]{Howlett2017}
{Howlett} C.,  {Staveley-Smith} L.,   {Blake} C.,  2017a, \mn@doi [\mnras]
  {10.1093/mnras/stw2466}, \href
  {https://ui.adsabs.harvard.edu/abs/2017MNRAS.464.2517H} {464, 2517}

\bibitem[\protect\citeauthoryear{{Howlett} et~al.,}{{Howlett}
  et~al.}{2017b}]{Howlett17}
{Howlett} C.,  et~al., 2017b, \mn@doi [\mnras] {10.1093/mnras/stx1521}, \href
  {https://ui.adsabs.harvard.edu/abs/2017MNRAS.471.3135H} {471, 3135}

\bibitem[\protect\citeauthoryear{{Hu} \& {White}}{{Hu} \& {White}}{1997}]{Hu97}
{Hu} W.,  {White} M.,  1997, \mn@doi [\prd] {10.1103/PhysRevD.56.596}, \href
  {https://ui.adsabs.harvard.edu/abs/1997PhRvD..56..596H} {56, 596}

\bibitem[\protect\citeauthoryear{{Hudson} \& {Turnbull}}{{Hudson} \&
  {Turnbull}}{2012}]{Hudson12}
{Hudson} M.~J.,  {Turnbull} S.~J.,  2012, \mn@doi [\apjl]
  {10.1088/2041-8205/751/2/L30}, \href
  {https://ui.adsabs.harvard.edu/abs/2012ApJ...751L..30H} {751, L30}

\bibitem[\protect\citeauthoryear{{Johnson} et~al.,}{{Johnson}
  et~al.}{2014}]{Johnson14}
{Johnson} A.,  et~al., 2014, \mn@doi [\mnras] {10.1093/mnras/stu1615}, \href
  {https://ui.adsabs.harvard.edu/abs/2014MNRAS.444.3926J} {444, 3926}

\bibitem[\protect\citeauthoryear{{Kaiser}}{{Kaiser}}{1987}]{Kai87}
{Kaiser} N.,  1987, \mn@doi [\mnras] {10.1093/mnras/227.1.1}, \href
  {http://adsabs.harvard.edu/abs/1987MNRAS.227....1K} {227, 1}

\bibitem[\protect\citeauthoryear{{Kim} et~al.,}{{Kim} et~al.}{2019}]{Kim19}
{Kim} A.,  et~al., 2019, \baas, \href
  {https://ui.adsabs.harvard.edu/abs/2019BAAS...51c.140K} {51, 140}

\bibitem[\protect\citeauthoryear{Koda et~al.,}{Koda et~al.}{2014}]{Koda2013}
Koda J.,  et~al., 2014, \mn@doi [Mon. Not. Roy. Astron. Soc.]
  {10.1093/mnras/stu1610}, 445, 4267

\bibitem[\protect\citeauthoryear{{Koribalski}}{{Koribalski}}{2012}]{Wallaby}
{Koribalski} B.~S.,  2012, \mn@doi [\pasa] {10.1071/AS12030}, \href
  {https://ui.adsabs.harvard.edu/abs/2012PASA...29..359K} {29, 359}

\bibitem[\protect\citeauthoryear{{Lahav}, {Fisher}, {Hoffman}, {Scharf}  \&
  {Zaroubi}}{{Lahav} et~al.}{1994}]{Lahav94}
{Lahav} O.,  {Fisher} K.~B.,  {Hoffman} Y.,  {Scharf} C.~A.,   {Zaroubi} S.,
  1994, \mn@doi [\apjl] {10.1086/187244}, \href
  {http://adsabs.harvard.edu/abs/1994ApJ...423L..93L} {423, L93}

\bibitem[\protect\citeauthoryear{{Lavaux} \& {Hudson}}{{Lavaux} \&
  {Hudson}}{2011}]{LavauxHudson}
{Lavaux} G.,  {Hudson} M.~J.,  2011, \mn@doi [\mnras]
  {10.1111/j.1365-2966.2011.19233.x}, \href
  {https://ui.adsabs.harvard.edu/abs/2011MNRAS.416.2840L} {416, 2840}

\bibitem[\protect\citeauthoryear{{Liu}, {Zhang}  \& {Parsons}}{{Liu}
  et~al.}{2016}]{Liu16}
{Liu} A.,  {Zhang} Y.,   {Parsons} A.~R.,  2016, \mn@doi [\apj]
  {10.3847/1538-4357/833/2/242}, \href
  {http://adsabs.harvard.edu/abs/2016ApJ...833..242L} {833, 242}

\bibitem[\protect\citeauthoryear{{Ma}, {Gordon}  \& {Feldman}}{{Ma}
  et~al.}{2011}]{Ma11}
{Ma} Y.-Z.,  {Gordon} C.,   {Feldman} H.~A.,  2011, \mn@doi [\prd]
  {10.1103/PhysRevD.83.103002}, \href
  {https://ui.adsabs.harvard.edu/abs/2011PhRvD..83j3002M} {83, 103002}

\bibitem[\protect\citeauthoryear{{Nicola}, {Refregier}, {Amara}  \&
  {Paranjape}}{{Nicola} et~al.}{2014}]{Nic14}
{Nicola} A.,  {Refregier} A.,  {Amara} A.,   {Paranjape} A.,  2014, \mn@doi
  [\prd] {10.1103/PhysRevD.90.063515}, \href
  {http://adsabs.harvard.edu/abs/2014PhRvD..90f3515N} {90, 063515}

\bibitem[\protect\citeauthoryear{{Nusser}}{{Nusser}}{2017}]{Nusser17}
{Nusser} A.,  2017, \mn@doi [\mnras] {10.1093/mnras/stx1225}, \href
  {https://ui.adsabs.harvard.edu/abs/2017MNRAS.470..445N} {470, 445}

\bibitem[\protect\citeauthoryear{{Padmanabhan}, {Tegmark}  \&
  {Hamilton}}{{Padmanabhan} et~al.}{2001}]{Pad01}
{Padmanabhan} N.,  {Tegmark} M.,   {Hamilton} A.~J.~S.,  2001, \mn@doi [\apj]
  {10.1086/319700}, \href {http://adsabs.harvard.edu/abs/2001ApJ...550...52P}
  {550, 52}

\bibitem[\protect\citeauthoryear{{Padmanabhan} et~al.,}{{Padmanabhan}
  et~al.}{2007}]{Pad07}
{Padmanabhan} N.,  et~al., 2007, \mn@doi [\mnras]
  {10.1111/j.1365-2966.2007.11593.x}, \href
  {http://adsabs.harvard.edu/abs/2007MNRAS.378..852P} {378, 852}

\bibitem[\protect\citeauthoryear{{Park}}{{Park}}{2000}]{Park00}
{Park} C.,  2000, \mn@doi [\mnras] {10.1046/j.1365-8711.2000.03886.x}, \href
  {https://ui.adsabs.harvard.edu/abs/2000MNRAS.319..573P} {319, 573}

\bibitem[\protect\citeauthoryear{{Passaglia}, {Manzotti}  \&
  {Dodelson}}{{Passaglia} et~al.}{2017}]{Pas17}
{Passaglia} S.,  {Manzotti} A.,   {Dodelson} S.,  2017, \mn@doi [\prd]
  {10.1103/PhysRevD.95.123508}, \href
  {http://adsabs.harvard.edu/abs/2017PhRvD..95l3508P} {95, 123508}

\bibitem[\protect\citeauthoryear{{Peacock}}{{Peacock}}{1999}]{Pea99}
{Peacock} J.~A.,  1999, {Cosmological Physics}

\bibitem[\protect\citeauthoryear{{Percival} et~al.,}{{Percival}
  et~al.}{2004}]{Percival04}
{Percival} W.~J.,  et~al., 2004, \mn@doi [\mnras]
  {10.1111/j.1365-2966.2004.08146.x}, \href
  {http://adsabs.harvard.edu/abs/2004MNRAS.353.1201P} {353, 1201}

\bibitem[\protect\citeauthoryear{{Pratten} \& {Munshi}}{{Pratten} \&
  {Munshi}}{2013}]{Pratten13}
{Pratten} G.,  {Munshi} D.,  2013, \mn@doi [\mnras] {10.1093/mnras/stt1854},
  \href {http://adsabs.harvard.edu/abs/2013MNRAS.436.3792P} {436, 3792}

\bibitem[\protect\citeauthoryear{{Qin}, {Howlett}  \& {Staveley-Smith}}{{Qin}
  et~al.}{2019}]{Qin19}
{Qin} F.,  {Howlett} C.,   {Staveley-Smith} L.,  2019, \mn@doi [\mnras]
  {10.1093/mnras/stz1576}, \href
  {https://ui.adsabs.harvard.edu/abs/2019MNRAS.487.5235Q} {487, 5235}

\bibitem[\protect\citeauthoryear{Raccanelli, Bertacca, Jeong, Neyrinck  \&
  Szalay}{Raccanelli et~al.}{2018}]{Raccanelli2016}
Raccanelli A.,  Bertacca D.,  Jeong D.,  Neyrinck M.~C.,   Szalay A.~S.,  2018,
  \mn@doi [Phys. Dark Univ.] {10.1016/j.dark.2017.12.003}, 19, 109

\bibitem[\protect\citeauthoryear{{Reimberg}, {Bernardeau}  \&
  {Pitrou}}{{Reimberg} et~al.}{2016}]{Rei16}
{Reimberg} P.,  {Bernardeau} F.,   {Pitrou} C.,  2016, \mn@doi [\jcap]
  {10.1088/1475-7516/2016/01/048}, \href
  {http://adsabs.harvard.edu/abs/2016JCAP...01..048R} {1, 048}

\bibitem[\protect\citeauthoryear{{Samushia}}{{Samushia}}{2019}]{Samushia19}
{Samushia} L.,  2019, arXiv e-prints, \href
  {https://ui.adsabs.harvard.edu/abs/2019arXiv190605866S} {p. arXiv:1906.05866}

\bibitem[\protect\citeauthoryear{{Scoccimarro}}{{Scoccimarro}}{2015}]{Sco15}
{Scoccimarro} R.,  2015, \mn@doi [\prd] {10.1103/PhysRevD.92.083532}, \href
  {http://adsabs.harvard.edu/abs/2015PhRvD..92h3532S} {92, 083532}

\bibitem[\protect\citeauthoryear{{Shaw}, {Sigurdson}, {Pen}, {Stebbins}  \&
  {Sitwell}}{{Shaw} et~al.}{2014}]{Sha14}
{Shaw} J.~R.,  {Sigurdson} K.,  {Pen} U.-L.,  {Stebbins} A.,   {Sitwell} M.,
  2014, \mn@doi [\apj] {10.1088/0004-637X/781/2/57}, \href
  {http://adsabs.harvard.edu/abs/2014ApJ...781...57S} {781, 57}

\bibitem[\protect\citeauthoryear{{Strauss} \& {Willick}}{{Strauss} \&
  {Willick}}{1995}]{SW95}
{Strauss} M.~A.,  {Willick} J.~A.,  1995, \mn@doi [\physrep]
  {10.1016/0370-1573(95)00013-7}, \href
  {https://ui.adsabs.harvard.edu/abs/1995PhR...261..271S} {261, 271}

\bibitem[\protect\citeauthoryear{{Szalay}, {Matsubara}  \& {Landy}}{{Szalay}
  et~al.}{1998}]{Sza98}
{Szalay} A.~S.,  {Matsubara} T.,   {Landy} S.~D.,  1998, \mn@doi [\apjl]
  {10.1086/311293}, \href
  {https://ui.adsabs.harvard.edu/abs/1998ApJ...498L...1S} {498, L1}

\bibitem[\protect\citeauthoryear{{Tadros} et~al.,}{{Tadros}
  et~al.}{1999}]{Tadros99}
{Tadros} H.,  et~al., 1999, \mn@doi [\mnras]
  {10.1046/j.1365-8711.1999.02409.x}, \href
  {http://adsabs.harvard.edu/abs/1999MNRAS.305..527T} {305, 527}

\bibitem[\protect\citeauthoryear{Taruya, Saga, Breton, Rasera  \&
  Fujita}{Taruya et~al.}{2019}]{Taruya}
Taruya A.,  Saga S.,  Breton M.-A.,  Rasera Y.,   Fujita T.,  2019

\bibitem[\protect\citeauthoryear{{Taylor}, {Ballinger}, {Heavens}  \&
  {Tadros}}{{Taylor} et~al.}{2001}]{Tay01}
{Taylor} A.~N.,  {Ballinger} W.~E.,  {Heavens} A.~F.,   {Tadros} H.,  2001,
  \mn@doi [\mnras] {10.1046/j.1365-8711.2001.04770.x}, \href
  {http://adsabs.harvard.edu/abs/2001MNRAS.327..689T} {327, 689}

\bibitem[\protect\citeauthoryear{{Tully}, {Courtois}  \& {Sorce}}{{Tully}
  et~al.}{2016}]{Tully16}
{Tully} R.~B.,  {Courtois} H.~M.,   {Sorce} J.~G.,  2016, \mn@doi [\aj]
  {10.3847/0004-6256/152/2/50}, \href
  {https://ui.adsabs.harvard.edu/abs/2016AJ....152...50T} {152, 50}

\bibitem[\protect\citeauthoryear{{Wang}, {Reid}  \& {White}}{{Wang}
  et~al.}{2014}]{Wang14}
{Wang} L.,  {Reid} B.,   {White} M.,  2014, \mn@doi [\mnras]
  {10.1093/mnras/stt1916}, \href
  {https://ui.adsabs.harvard.edu/abs/2014MNRAS.437..588W} {437, 588}

\bibitem[\protect\citeauthoryear{{Weinberg}, {Mortonson}, {Eisenstein},
  {Hirata}, {Riess}  \& {Rozo}}{{Weinberg} et~al.}{2013}]{Weinberg13}
{Weinberg} D.~H.,  {Mortonson} M.~J.,  {Eisenstein} D.~J.,  {Hirata} C.,
  {Riess} A.~G.,   {Rozo} E.,  2013, \mn@doi [\physrep]
  {10.1016/j.physrep.2013.05.001}, \href
  {http://adsabs.harvard.edu/abs/2013PhR...530...87W} {530, 87}

\bibitem[\protect\citeauthoryear{{Yamamoto}, {Nakamichi}, {Kamino}, {Bassett}
  \& {Nishioka}}{{Yamamoto} et~al.}{2006}]{Yam06}
{Yamamoto} K.,  {Nakamichi} M.,  {Kamino} A.,  {Bassett} B.~A.,   {Nishioka}
  H.,  2006, \mn@doi [\pasj] {10.1093/pasj/58.1.93}, \href
  {http://adsabs.harvard.edu/abs/2006PASJ...58...93Y} {58, 93}

\bibitem[\protect\citeauthoryear{{Yoo} \& {Desjacques}}{{Yoo} \&
  {Desjacques}}{2013}]{YD2013}
{Yoo} J.,  {Desjacques} V.,  2013, \mn@doi [\prd] {10.1103/PhysRevD.88.023502},
  \href {http://adsabs.harvard.edu/abs/2013PhRvD..88b3502Y} {88, 023502}

\bibitem[\protect\citeauthoryear{{Zaroubi} \& {Hoffman}}{{Zaroubi} \&
  {Hoffman}}{1996}]{ZH96}
{Zaroubi} S.,  {Hoffman} Y.,  1996, \mn@doi [\apj] {10.1086/177124}, \href
  {http://adsabs.harvard.edu/abs/1996ApJ...462...25Z} {462, 25}

\bibitem[\protect\citeauthoryear{{Zel'dovich}}{{Zel'dovich}}{1970}]{Zel70}
{Zel'dovich} Y.~B.,  1970, \aap, \href
  {http://adsabs.harvard.edu/abs/1970A%26A.....5...84Z} {5, 84}

\bibitem[\protect\citeauthoryear{{da Cunha} et~al.,}{{da Cunha}
  et~al.}{2017}]{Taipan}
{da Cunha} E.,  et~al., 2017, \mn@doi [\pasa] {10.1017/pasa.2017.41}, \href
  {https://ui.adsabs.harvard.edu/abs/2017PASA...34...47D} {34, e047}

\makeatother
\end{thebibliography}
\end{document}